\documentclass{aastex}
\usepackage{spr-astr-addons}
\usepackage{url}

\RequirePackage{color}

\onecolumn

\begin{document}

\title{ACRIM total solar irradiance satellite composite validation versus
TSI proxy models}
\shorttitle{ACRIM TSI satellite composite versus proxy models}
\shortauthors{N. Scafetta and R. C. Willson}

\author{Nicola Scafetta\altaffilmark{1,2}} \and \author{Richard C. Willson\altaffilmark{1}}

\email{nicola.scafetta@gmail.com}

\altaffiltext{1}{1Active Cavity Radiometer Irradiance Monitor (ACRIM) Lab, Coronado, CA 92118, USA.}
\altaffiltext{2}{Duke University}

\onecolumn

\begin{abstract}
The satellite total solar irradiance (TSI) database provides a valuable
record for investigating models of solar variation used to interpret
climate changes. The 35-year ACRIM total solar irradiance (TSI) satellite
composite time series has been updated using corrections to ACRIMSAT/ACRIM3
results for scattering and diffraction derived from recent testing
at the Laboratory for Atmospheric and Space Physics/Total solar irradiance
Radiometer Facility (LASP/TRF). The corrections lower the ACRIM3 scale
by about 5000 ppm, in close agreement with the scale of
SORCE/TIM results (solar constant $\thickapprox1361$ $W/m^{2}$)
but the relative variations and trends are not changed. Differences
between the ACRIM and PMOD TSI composites, particularly the decadal
trending during solar cycles 21-22, are tested against a set of solar
proxy models, including analysis of Nimbus7/ERB and ERBS/ERBE results
available to bridge the ACRIM Gap (1989-1992). Our findings confirm
the following ACRIM TSI composite features: (1) The validity of the
TSI peak in the originally published ERB results in early 1979 during
solar cycle 21; (2) The correctness of originally published ACRIM1
results during the SMM spin mode (1981\textendash{}1984); (3) The
upward trend of originally published ERB results during the ACRIM
Gap; (4) The occurrence of a significant upward TSI trend between
the minima of solar cycles 21 and 22 and (5) a decreasing trend during
solar cycles 22 - 23. Our findings do not support the following PMOD
TSI composite features: (1) The downward corrections to originally
published ERB and ACRIM1 results during solar cycle 21; (2) A step
function sensitivity change in ERB results at the end-of-September
1989; (3) the validity of ERBE\textquoteright{}s downward trend during
the ACRIM Gap or (4) the use of ERBE results to bridge the ACRIM Gap.
Our analysis provides a first order validation of the ACRIM TSI composite
approach and its 0.037\%/decade upward trend during solar cycles 21-22.
The implications of increasing TSI during the global warming of the
last two decades of the 20th century are that solar forcing of climate
change may be a significantly larger factor than represented in the
CMIP5 general circulation climate models.\\ {} \\ \textbf{Cite:} Scafetta, N., and R. C. Willson, 2014. ACRIM total solar irradiance satellite composite validation versus TSI proxy models. Astrophysics and Space Science 350(2), 421-442.
DOI: 10.1007/s10509-013-1775-9.
\end{abstract}


\twocolumn

\section{Introduction}

The satellite total solar irradiance (TSI) database is now more than
three and a half decades long and provides a valuable record for investigating
the relative significance of natural and anthropogenic forcing of
climate change \citep{IPCC,Scafetta2009,Scafetta2011}. It is made
of 7 major independent measurements covering different periods since
1978 (see Figure 1).

\begin{figure*}[!t]
\centering
\includegraphics[width=1.8\columnwidth]{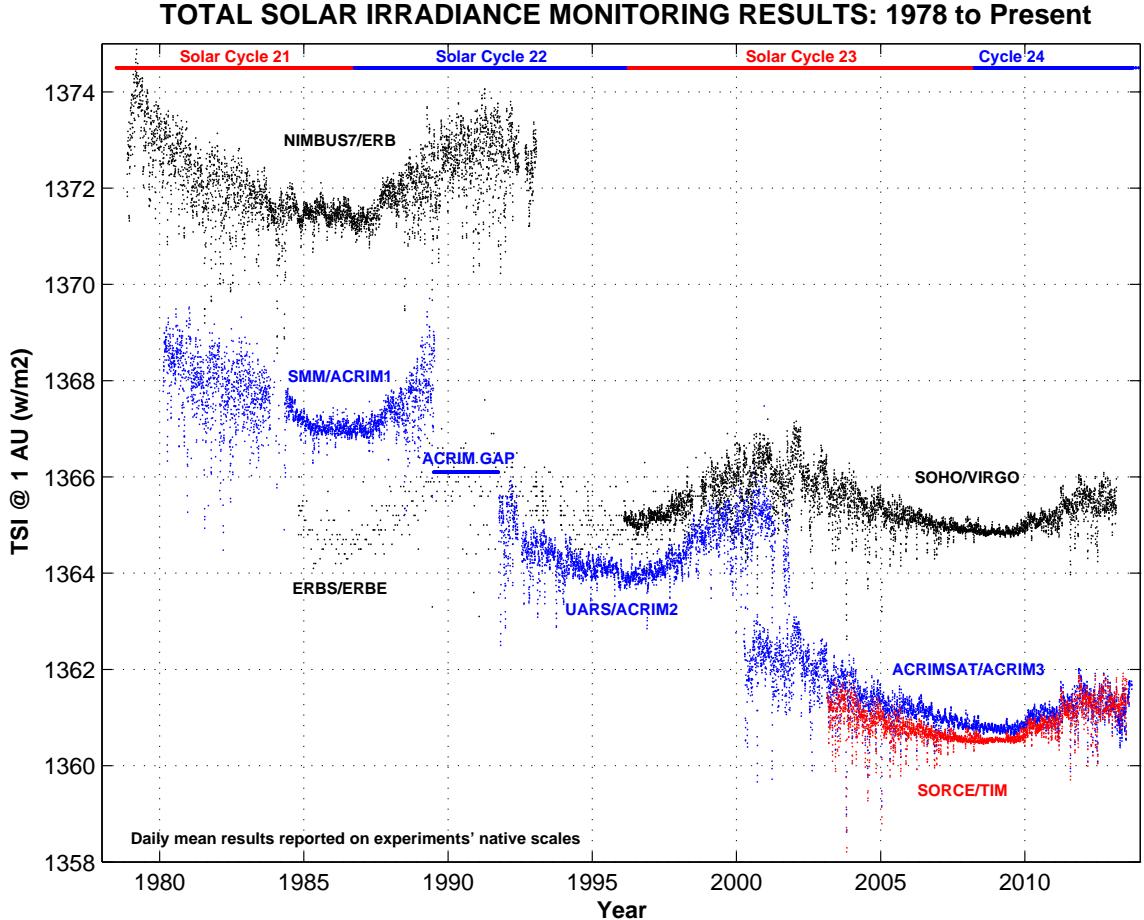}
\caption{Total solar irradiance satellite record database. }
\end{figure*}

A composite TSI record can be constructed from the series of experiments
since 1978 by combining and cross-calibrating the set of overlapping
satellite observations to create a TSI time series. TSI satellite
composites provide end-to-end traceability at the mutual precision
level of the overlapping satellite experiments that is orders of magnitude
smaller than the absolute uncertainty of the individual experiments.
The scale offsets of the various satellite results shown in Figure
1 are caused by the uncertainties of their self-calibration \citep{Willson2003,Frohlich2012}.
Different approaches in selecting results and cross-calibrating the
satellite records on a common scale have resulted in composites with
different characteristics.

\begin{figure*}[!t]
\centering
\includegraphics[width=1.8\columnwidth]{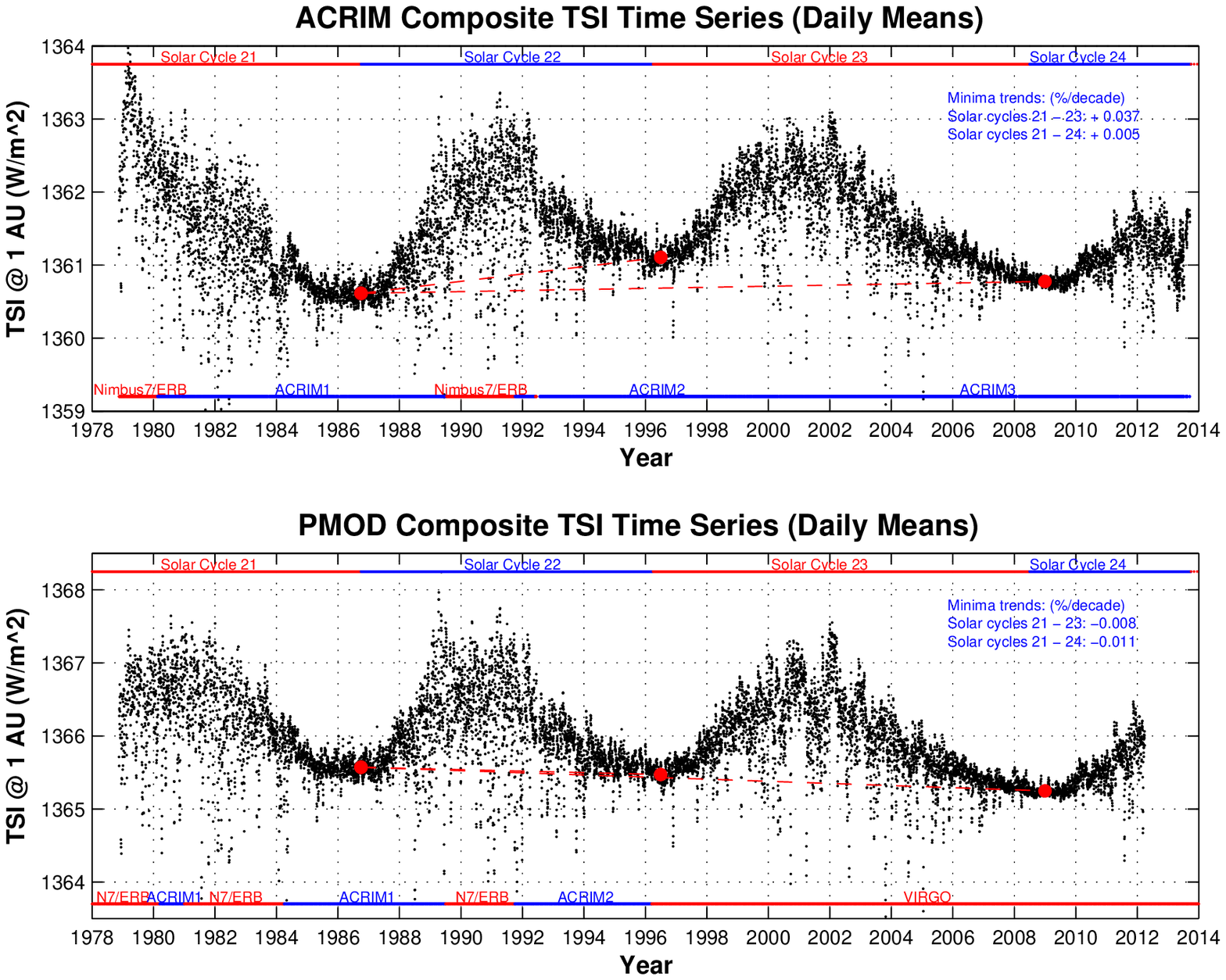}\caption{TSI satellite composites ACRIM and PMOD. Different approaches to bridging
the ACRIM Gap result in different trends. The ACRIM composite uses:
(1) ERB, ACRIM1, 2 and 3 results published by the experiment science
teams; (2) ERB comparisons to bridge the ACRIM Gap; (3) ACRIM3 scale.
The PMOD composite uses: ERB, ACRIM1, ACRIM2 and VIRGO results; (2)
ERBE comparisons to bridge the ACRIM Gap; (3) Alters published ERB
and ACRIM1 results to conform them to TSI proxy models; (4) VIRGO
scale. }
\end{figure*}

Figure 2 shows the two TSI satellite composites most commonly cited:
ACRIM \citep{Willson1997,Willson2001,Willson2003} and PMOD \citep{Frohlich1998,Frohlich2004,Frohlich2006,Frohlich2012}.
Alternative TSI satellite composites have been proposed by \citet{Dewitte}
and \citet{Scafetta2011} using different methodologies to merge the
datasets.

The new ACRIM composite uses the updated  ACRI M3 record. ACRIM3 data
was reprocessed after implementing corrections for scattering and
diffraction found during recent testing and some other algorithm updates.
The testing was performed at the TSI Radiation Facility (TRF) of the
Laboratory for Atmospheric and Space Physics (LASP) \citep[\url{http://lasp.colorado.edu/home/}]{Kopp}.
Two additional algorithm updates were implemented that more accurately
account for instrument thermal behavior and parsing of shutter cycle
data. These removed a component of the quasi-annual signal from the
data and increased the signal to noise ratio of the data, respectively.
The net effect of these corrections decreased the average ACRIM3 TSI
value from $\sim1366$ $W/m^{2}$ \citep[see: ][]{Willson2003} to
$\sim1361$ $W/m^{2}$ without affecting the trending in the ACRIM
Composite TSI.

Differences between ACRIM and PMOD TSI composites are evident, but
the most obvious and significant one is the solar minimum-to-minimum
trends during solar cycles 21 to 23. ACRIM presents a bi-decadal increase
of +0.037\%/decade from 1980 to 2000 and a decrease thereafter. PMOD
presents a steady multi-decadal decrease since 1978 (see Figure 2).
Other significant differences can be seen during the peak of solar
cycles 21 and 22. These arise from the fact that ACRIM uses the original
TSI results published by the satellite experiment teams while PMOD
significantly modifies some results to conform them to specific TSI
proxy models \citep{Frohlich1998,Frohlich2004,Frohlich2006,Frohlich2012}.

The single greatest challenge in constructing a precise composite
extending before 1991 is providing continuity across the two-year
ACRIM Gap (1989.53\textendash{} 1991.76) between the results of SMM/ACRIM1
\citep{Willson1991} and UARS/ACRIM2 \citep{Willson1994,Willson1997}.
During this period the only observations available were those of the
Nimbus7/ERB (hereafter referred to as ERB) \citep{Hoyt1992} and ERBS/ERBE
(hereafter referred to as ERBE) \citep{Lee}. These experiments provided
TSI observations that met the needs of the Earth Radiation Budget
investigations at that time, but were less precise and accurate than
the ACRIM experiments that were designed specifically to provide the
long term precision and traceability required by climate and solar
physics investigations.

ACRIM1 and ACRIM2 were intended to overlap initiating an ACRIM TSI
monitoring strategy designed to provide long term TSI traceability
of results through the precision of on-orbit comparisons. ACRIM2 was
delayed by the Challenger disaster, however, and eventually deployed
two years after the last data from ACRIM1. This period is known as
the ACRIM GAP (1989.5 - 1991.75), as shown in Figure 1.

ACRIM1, ACRIM2 and ACRIM3 were dedicated TSI monitoring experiments
capable of highly precise observations by virtue of their design and
operation, which includes continuous electronic self-calibration,
high duty cycle solar observations (ACRIM1: 55 min./orbit; ACRIM2:
35 min./orbit; ACRIM3: up to full sun during its 96 minute sun-synchronous
orbit), sensor degradation self-calibration, high observational cadence
($~2$ minutes) and precise solar pointing. ERB and ERBE were less
accurate and precise experiments designed to meet the less stringent
data requirements of Earth Radiation Budget modeling. They were able
to self-calibrate only infrequently (every 14 days), had limited solar
observational opportunities (ERB: 5 min/orbit daily; ERBE: 5 minutes
every 14 days, usually) and were not independently solar pointed,
observing while the sun moved through their fields of view, all of
which degraded their precision and accuracy.

Bridging the ACRIM Gap using ERB and ERBE results is problematical
not only because of their lower data quality but also because their
results yield significantly different and incompatible trends during
the ACRIM Gap. During the ACRIM Gap ERB results trend upward (linear
regression slope = $0.27\pm0.04$ $Wm^{-2}/year$) while ERBE trend
downward (linear regression slope =$-0.27\pm0.15$ $Wm^{-2}/year$).
This causes the difference between the ACRIM and PMOD TSI trends during
solar cycles 21-23. The ACRIM TSI composite uses unaltered ERB results
to relate ACRIM1 and ACRIM2 records, while PMOD uses an altered ERB
record based on some theoretical model predictions that better agree
with the downward trend of the ERBE record during the ACRIM Gap.

\begin{figure*}[!t]
\centering
\includegraphics[width=1.4\columnwidth, angle=-90]{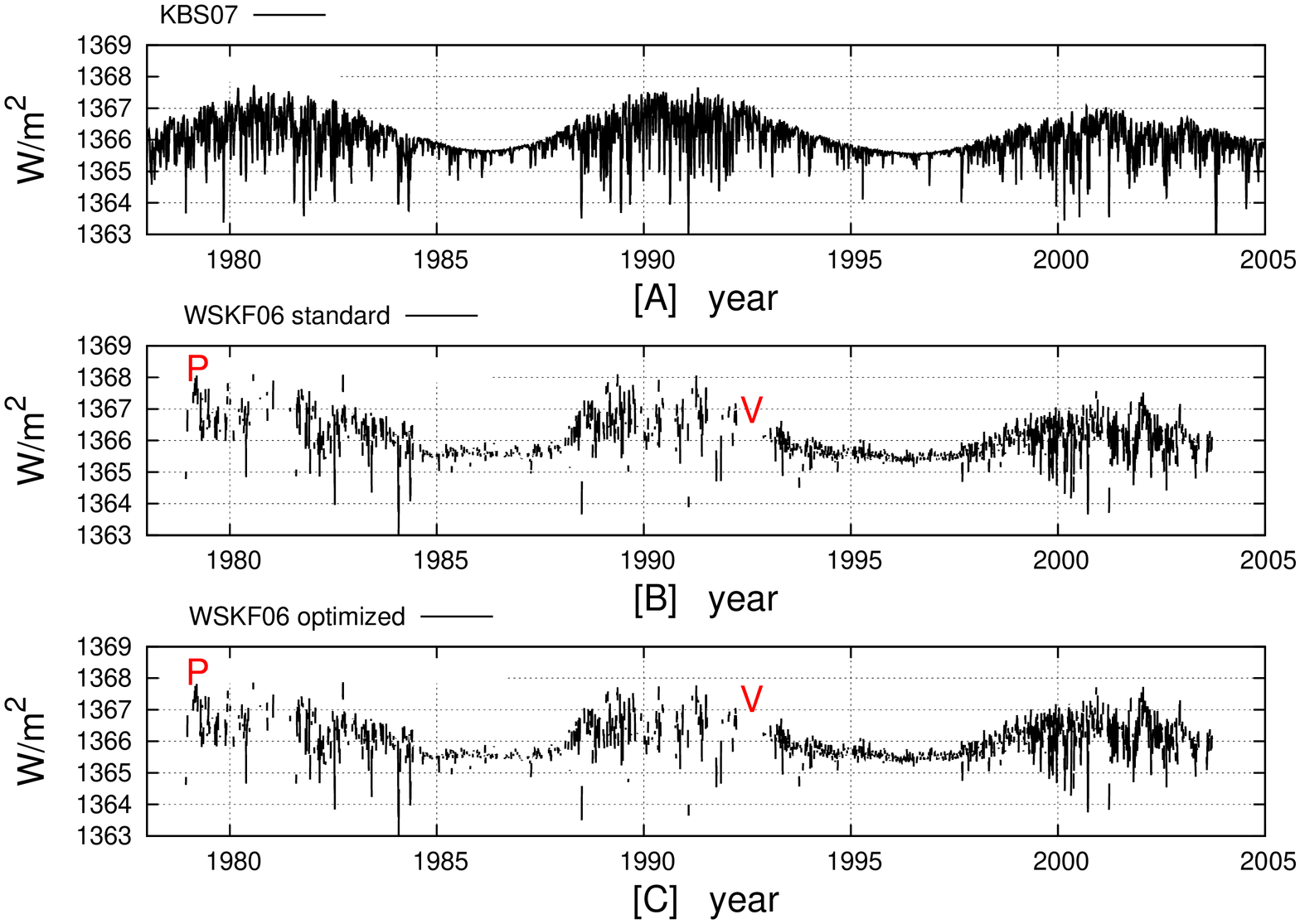}\caption{{[}A{]} The \citet{Krivova2007} magnetic field proxy model. {[}B{]}
\citet{Wenzler2006} (WSKF06) standard surface magnetic field proxy
models. {[}C{]} \citet{Wenzler2006} (WSKF06) optimized (on PMOD)
standard surface magnetic field proxy models. The 1979 peak maximum
(red letter ``P'') in the WSKF06 models and the lack of available
data in 1992 (red letter \textquotedblleft{}V\textquotedblright{})
that separates NSO-512 (Feb/1/74 to Apr/18/92 and Nov/28/92 to Apr/10/93)
and NSO-SPM (Nov/21/92 to Sep/21/03) records calls into question the
accuracy of their cross-calibration with the KBS07 model. }
\end{figure*}

In Section 2 we review the hypotheses proposed in the literature about
Nimbus7/ERB TSI record during the ACRIM Gap. In sections 3-8 we test
these hypotheses by directly comparing the Nimbus7/ERB data sets versus
alternative solar data and proxy models. In this process we will study
in details the TSI proxy models of \citet{Krivova2007} (KBS07), and
\citet{Wenzler2006} (WSKF06) shown in Figure 3. In Appendix A Hoyt
(the head of the NASA Nimbus7/ERB science team) explains the accuracy
of the ERB record during the ACRIM Gap. Appendix B briefly summarizes
the importance of the TSI satellite composite issue for solar physics
and climate change.

\section{Review of the PMOD hypotheses about Nimbus7/ERB TSI record}

The PMOD composite is constructed using ERB, ACRIM1, ERBE, ACRIM2
and VIRGO results. Some ERB and ACRIM1 published results were modified
in the process \citep{Frohlich1998,Frohlich2004}. These modifications
were not based on re-analysis of satellite instrumentation or data
but on an effort to conform the satellite TSI record to the predictions
of TSI proxy models developed by \citet{Lean} and \citet{Lee}. These
proxy models are statistical regressions containing no physics and
cannot be considered to be competitive in accuracy or precision with
the satellite TSI observations themselves.

More recently \citet{Frohlich2006} endeavored to justify alteration
of ERB results during its early mission using a theoretical model
based on the initial on-orbit degradation of VIRGO TSI sensors and
the similarity of VIRGO and ERB sensors. However, this approach cannot
be used to justify the ERB modifications. It\textquoteright{}s well
known that on-orbit exposure to solar fluxes is the principal cause
of sensor degradation and solar exposure was radically different for
the VIRGO and ERB missions. VIRGO received constant solar irradiation
at its L1 solar orbit because there is no Earth shadow and the shutters
for the sensors failed during launch. Irradiation of ERB sensors was
far less with solar exposure only 5 minutes each orbit during the
three days out of every four it operated in the Nimbus7 Earth orbit.The
absorbent surface coating of the VIRGO and ERB sensors exhibited very
different degradation with the VIRGO sensors showing the largest degradation
($\sim$5000 ppm over the mission) yet seen in TSI satellite
instrumentation. ERB degradation was more than an order of magnitude
less as its results tracked the ACRIM1 experiment during nearly a
decade.

The most controversial modification of published ERB results by the
PMOD composite was the assignment of a sensitivity shift during the
ACRIM Gap. The shift made the ERB \textquoteleft{}gap\textquoteright{}
results agree in scale with the ERBE \textquoteleft{}gap\textquoteright{}
results and caused the decadal TSI trending to agree with the TSI
proxy models developed by Lean and Lee III. This was done without
any analysis of instrument performance, algorithm update or data processing.
It is noteworthy that the scientists most familiar with the ERB experiment
and its data, the instrument developer and Principal Investigator,
John Hickey, and the NASA ERB science team principal investigator,
Douglas Hoyt, both reject the sensitivity increases proposed in Lee
and Fr\"ohlich (see Appendix A for a statement written by Hoyt).

Let us review the various attempts to reconcile the ERB and ERBE results
during the ACRIM Gap using theoretical models. These have been quite
contradictory and deserve special attention.
\begin{enumerate}
\item \citet{Lee} hypothetized that ERB sensors experienced uncorrected
sensitivity increases during the ACRIM Gap using the predictions of
a simple TSI proxy model regressing the photometric sunspot index
(PSI) and the 10.7-cm solar radio flux (F10) against the satellite
TSI observations. Lee\textquoteright{}s model diverges from ERB results
after September 1989 while approximately reproducing the ERBE results.
The proxy model\textquoteright{}s indication of 0.03\% ERB sensitivity
increases during both September/1989 and April/1990 was Lee\textquoteright{}s
rationale for shifting ERB results during the ACRIM Gap downward by
0.06\% to agree with ERBE results.
\item \citet{Chapmam} developed a TSI proxy model that indicated ERB sensitivity
upward shifts of 0.02\% and 0.03\% in October/1989 and May/1990. They
shifted ERB results downward by 0.05\% to provide better agreement
between ERB and ERBE results during the ACRIM Gap.
\item \citet{Frohlich1998} indicated two ERB upward \textit{glitches} in
sensitivity occurring exactly on October/1/1989 and May/8/1990). A
total downward shift of 0.05\% was used to reconcile ERB to ERBE results
during the ACRIM Gap.
\item \citet{Frohlich2004} compared ERB and ERBE results and concluded
that ERB experienced a step sensitivity increase of 0.03\% on September/29/1989
and a continuing gradual upward linear drift between October/1989
and June/1992. Thus, ERB results had to be first shifted downward
on September/29/1989 and then inclined downward until June/1992. The
combined effects reconciled ERB and ERBE ACRIM Gap results.
\item \citet{Frohlich2006} developed another proxy model calibrated against
his corrected version of ACRIM1 and ERBE results. He proposed that
the similarity of ERB and VIRGO sensors would allow degradation analyses
for VIRGO to be applied to ERB results. This approach is called into
question by the fact that the VIRGO sensor has shown unusually large
degradation ($\sim$5000 ppm) during its mission and there is no evidence
ERB experienced a similar effect of comparable magnitude. Lastly,
degradation is directly tied to the amount of exposure to the sun
and this is very different for the ERBE and VIRGO PMO6-A sensor.
\end{enumerate}

\section{Direct analysis of PMOD hypothesis of an ERB sensitivity increase
during the ACRIM Gap}

\begin{figure*}[!t]
\centering
\includegraphics[width=1.4\columnwidth, angle=-90]{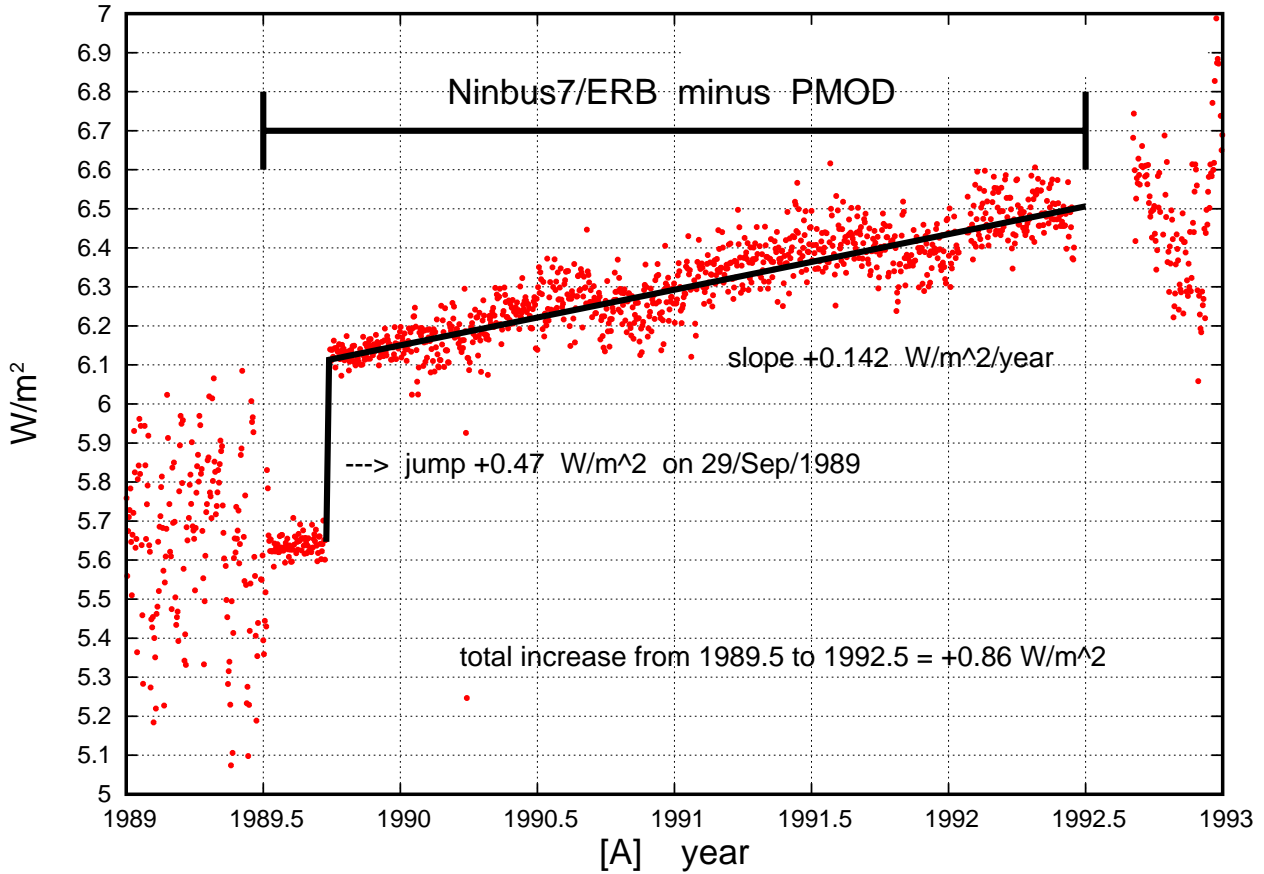}\caption{The PMOD corrections to the ERB record during the ACRIM Gap (1989.5-1992.5)
shown as the residual function Eq. \ref{eq:1} between Nimbus7/ERB
results and the PMOD TSI composite model.}
\end{figure*}

Figure 4 shows our analysis of the latest \citet{Frohlich2006} revision
of ERB results. The difference between published ERB results and the
PMOD composite during the ACRIM Gap is minimized by a step function
sensitivity change of +0.034\% (0.47 $W/m^{2}$) on September/29/1989
followed by a linear upward drift of 0.01\%/year from October/1989
through mid-1992. During the ACRIM Gap \citet{Frohlich2006} corrects
the ERB results by about 0.063\% (0.86 $W/m^{2}$), shifting them
downward to agree with ERBE \citep[see the detailed discusion in ][]{Scafetta2011}.
This corresponded also to the predictions of Lean's TSI proxy model
used for the previous version of the PMOD composite \citep{Frohlich1998,Frohlich2004}.
The direct consequence of Fr\"ohlich\textquoteright{}s revision is
that the PMOD TSI composite shows no significant trending between
the 1986 and 1996 solar minima.

Fr\"ohlich did not make original computations using ERB flight data
or indicate the statistical  uncertainty associated with his alteration
of the TSI experimental results originally published by the ERB science
team. The PMOD TSI composite data \footnote{\url{ftp://ftp.pmodwrc.ch/pub/data/irradiance/composite/DataPlots/}}
  reports TSI daily values with four decimal digits of precision. The
TSI values of the original satellite records have, on average, only
two decimal digits of precision. Clearly PMOD misrepresents the statistical
significance of its data which calls into question the validity of
PMOD trending.

\section{ACRIM Gap and the line of sight solar magnetic field strength (SMFS)
measurements}

A first order resolution of the ACRIM Gap issue can be made by observing
that the ERB TSI increase during the gap conforms to the Solar Magnetic
Activity/TSI (SMA/TSI) paradigm; a positive correlation between TSI
and solar magnetic activity discovered by satellite experiments during
the 1980\textquoteright{}s \citep{Willson1991,Willson1997}. The SMA/TSI
paradigm holds on time scales longer than a solar rotation and approximately
correlates with the quasi 11-year TSI cycle. It has been validated
by all experimental components of the set of TSI satellite monitoring
observations to date.

\begin{figure*}[!t]
\centering
\includegraphics[width=1.6\columnwidth]{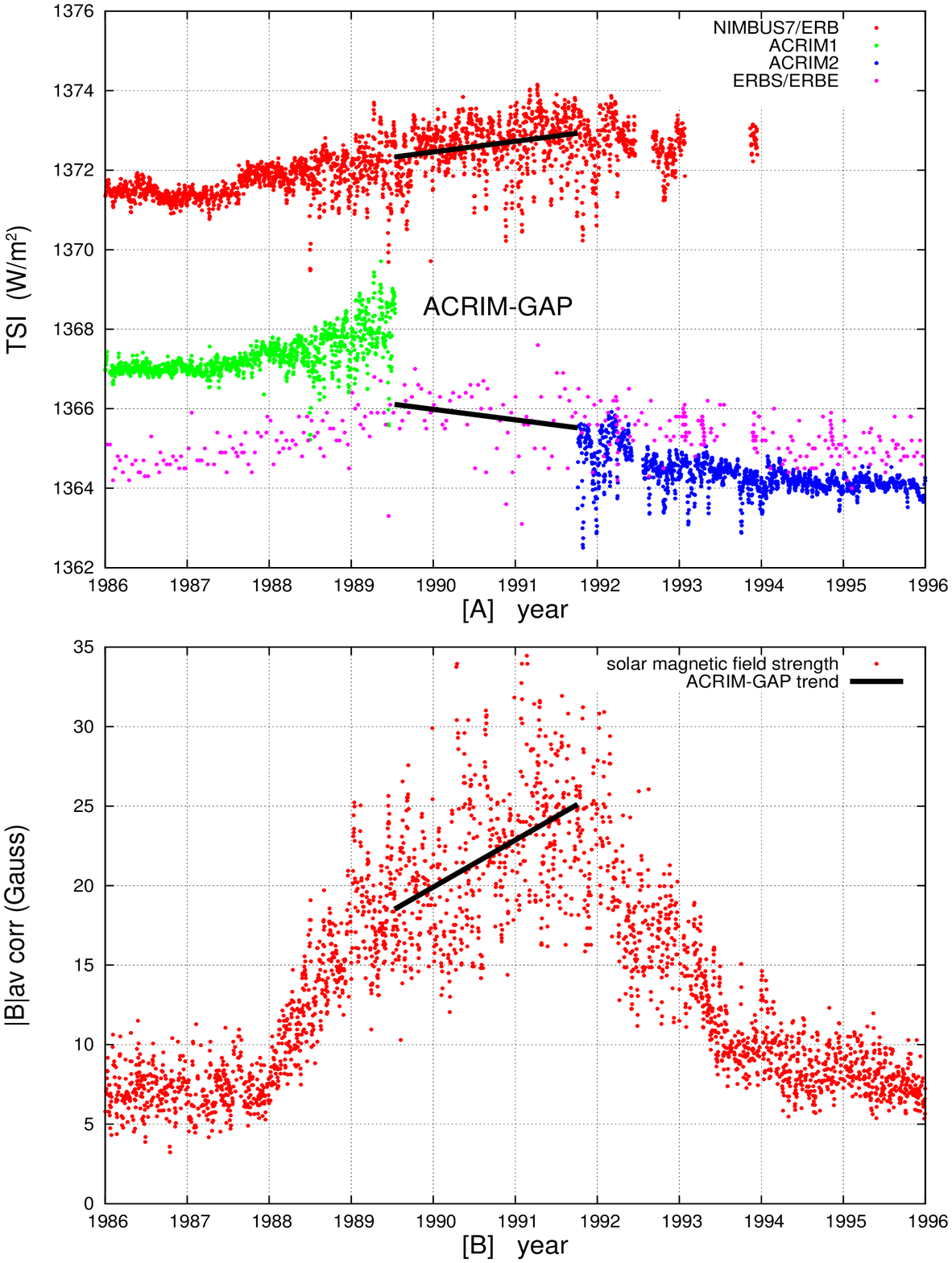}\caption{{[}A{]} ACRIM1, ACRIM2, Nimbus7/ERB and ERBS/ERBE original records
during solar cycle 22. {[}B{]} The solar magnetic field strength (National
Solar Observatory/Kitt Peak Data Archives: \protect\url{ftp://nsokp.nso.edu/kpvt/daily/stats/mag.dat}).
During the ACRIM Gap (1989.53-1991.76) the data clearly show an upward
trend (black segment) in the solar magnetic index (linear regression
slope = $3.0\pm0.3$ $Gauss/year$) that is consistent with the TSI
upward trend of Nimbus7/ERB record (linear regression slope = $0.27\pm0.04$
$Wm^{-2}/year$) but not with the TSI downward trend of ERBS/ERBE
record (linear regression slope =$-0.27\pm0.15$ $Wm^{-2}/year$). }
\end{figure*}

The SMA/TSI paradigm would be confirmed by the upward trend of the
daily line of sight solar magnetic field strength (SMFS) measurements
shown in Figure 5B during the ACRIM Gap. During the ACRIM Gap ERB
results trend upward with the SMFS record  and conform to the SMFS/TSI
paradigm. ERBE are anti-correlated, trending downward (compare Figures
5A and 5B). ERB results should therefore be considered the most likely
correct representation of TSI trending during the ACRIM Gap. Consequently,
the ACRIM TSI composite is the most likely correct representation
of the decadal TSI trend during solar cycles 21-23.

The most probable explanation for the ERBE negative trend correlation
with the SMFS during the ACRIM Gap could be uncorrected sensor degradation.
Rapid degradation of TSI sensors is commonly observed during initial
exposures to the enhanced UV fluxes that occur during periods of maximum
solar magnetic activity \citep{Willson2003}. During solar maxima
extreme UV photon energy flux is about two times larger than during
solar cycle minima \citep{Lean2005} and sensor degradation can occur
much faster.

Sensor degradation on most experiments saturates eventually, becoming
asymptotic after prolonged exposure to solar UV fluxes. The high and
rising level of solar magnetic activity during solar cycle 22 that
occurred during the ACRIM Gap was the first exposure of ERBE to enhanced
UV radiation. The solar cycle 22 maximum was the second exposure for
ERB, whose mission began in 1978 just before the maximum activity
period of solar cycle 21, and whose degradation had likely already
reached or was approaching its asymptote after 1985.

We note that TSI proxies, such as the sunspot number, the F10.7 radio
flux, and the Ca-II, Mg-II and He-I chromospheric lines that address
certain features and wavelength regions of the solar spectrum, do
not show a consistent upward trend during the ACRIM Gap \citep[figures 12 and 13]{Fox2004}.
However, SMFS, the solar magnetic flux index, should provide a more
robust and specific proxy for solar magnetic activity and TSI. Addressing
the ACRIM GAP TSI trending issue using solar proxies that address
specific wavelength ranges is a controversial approach. A more detailed
investigation using TSI data and solar magnetic activity proxy data
and models is required, and this is our approach.

\section{The Sep/29/1989 Nimbus7/ERB `glitch' hypothesis}

We will now examine the ERB `glitch' hypothesis of \citet{Lee} and
\citet{Frohlich2006} hypothesized to have occurred on Sep/29/1989.
Their ERB \textquoteleft{}glitch\textquoteright{} was derived using
a direct comparison of ERB record and of a proxy model calibrated
on ERBE records. Their claim is that ERB sensors exhibited a $\sim$0.4
$W/m^{2}$ sudden increase of sensitivity following a three-day instrument
power down (Sep. 25-28, 1989), implying that sensor properties changed.
Figure 6 shows the difference between ERB and ERBE results for days
between Jan/01/1989 and Jun/30/1990 when both experiments had solar
observations. The relevant TSI values are reported in Table 1.

\begin{figure*}[!t]
\centering
\includegraphics[width=1.4\columnwidth, angle=-90]{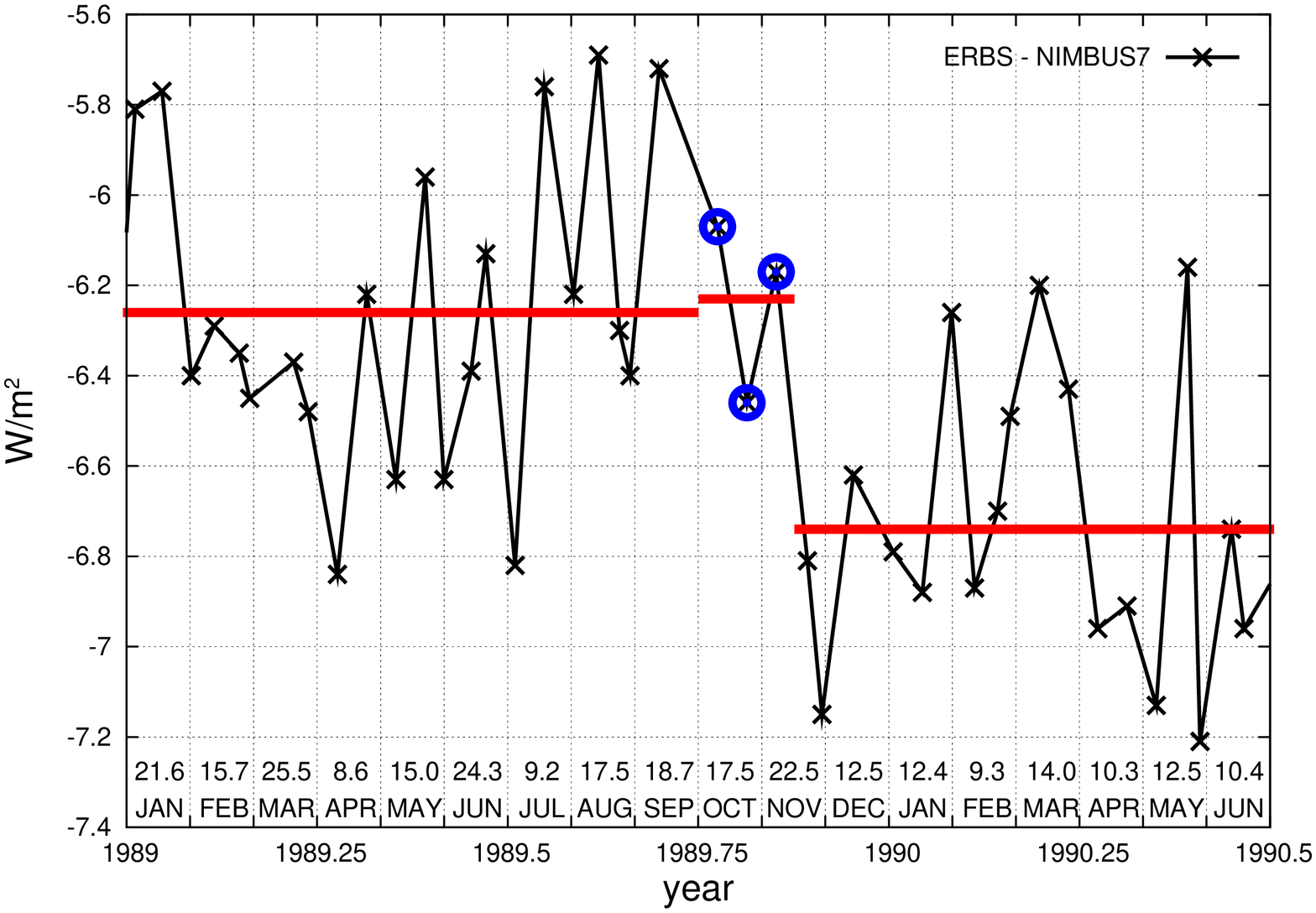}\caption{Residual function Eq. \ref{eq:1}between ERBS/ERBE and Nimbus7/ERB
results: see Table 1. The two records diverge mostly in November,
not at the end of September. The blue circle events in October and
November 1989 are statistically compatible with the values before
the end-of-September ERB \textquoteleft{}power-down\textquoteright{}
event. The red segments are mean values during the corresponding periods.
At the bottom of the figure the solar flare index (SFI) activity is
reported showing a peak during Nov/1989: SFI data from \protect\url{ftp://ftp.ngdc.noaa.gov/STP/space-weather/solar-data/solar-features/solar-flares/index/comprehensive-flare-index/documentation/cfi\_{}monthly\_{}1966-2008.txt}.}
\end{figure*}

Figure 6 clearly shows that ERB and ERBE diverge by $\sim0.5$ $W/m^{2}$
over this 1.5 year period. This divergence did not occur as a step
function during the three-day power-down of September 25-28, as postulated
by Lee and Fr\"ohlich. During the period Jan/1989 to the end of Sep/1989
the mean of the difference between the two records is $-6.26\pm0.35$
$W/m^{2}$. During the period Nov/23/1989 to Jun/20/1990 the mean
of the difference between the two records is $-6.74\pm0.32$ $W/m^{2}$.
The values on Oct/11, Oct/25 and Nov/08 statistically agree with the
previous Jan/1989-Set/1989 trend 23 times better than with the ensuing
Nov/23/1989-Jun/20/1990 trend: their mean value is $-6.23\pm0.20$
$W/m^{2}$. See also the discussion made in \citet[figure 7]{Scafetta2011}
that also highlights a rapid, but gradual divergence between Ninbus7/ERB
and ERBS/ERBE during October and November 1989 that contrasts with
the sudden one-day glitch shift claimed by PMOD on September 29, 1989.

Thus, the experimental evidence indicates that ERB and ERBE results
changed in November 1989, not at the end of September. This result
questions the single most important PMOD composite assumption that
produces different 1980-2000 trending from the ACRIM composite. It
also supports Hoyt\textquoteright{}s statement (see the Appendix):
\textit{``The calibrations before and after the September shutdown
gave no indication of any change in the sensitivity of the radiometer.''}

An ERBE sensitivity reduction at about that time may have had numerous
causes. For example, during 1989 there was an exceptionally high rate
of increase of solar activity due to the fact the Sun was approaching
the maximum of solar cycle 22. The ERBE-ERB divergence occurred during
November 1989 and this period coincided, for example, with an exceptionally
high value of the solar flare index (SFI) whose monthly means are
indicate in the bottom of Figure 6. The SFI maximum of the year occurred
on Oct/19/1989 (SFI = 89) and the SFI average for November was higher
(mean SFI = 22.5) than the previous months. Thus, the ERB-ERBE divergence
during that period could have been caused by a rapid degradation of
ERBE sensors by enhanced UV solar fluxes as the Sun entered in its
maximum level of activity. As explained in Sections 4 and 5, rapid
TSI sensor degradation occurs during their first exposure to the enhanced
short wavelength fluxes of solar maximum periods. This was the first
solar maximum experienced by ERBE but the second by ERB, which likely
had reached or was near its asymptotic degradation level. Solar pointing
issues could also have occurred.

There may be other physical explanations for the ERB-ERBE divergence
during the ACRIM Gap as discussed in section 4. However the Sep/1989
Nimbus7/ERB three-day `glitch' hypothesis proposed by Lee and Fr\"ohlich
is not supported by the experimental evidence reported in Figure 6.

\begin{table*}[!t]
\centering{}\begin{tabular}{|c|c|c|c|c|c|c|c|c|}\hline
date	&	ERBE 	&	ERB	&	diff	&	&	date	&	ERBE	&	ERB	&	diff	\\
	&	$W/m^2$ 	&	$W/m^2$	&	$W/m^2$	&	&		&	$W/m^2$	&	$W/m^2$	&	$W/m^2$	\\ \hline \hline
89/01/05	&	1366.4	&	1372.21	&	-5.81	&	&	89/10/11	&	1367.0	&	1373.07	&	-6.07	\\ \hline
89/01/18	&	1366.0	&	1371.77	&	-5.77	&	&	89/10/25	&	1366.6	&	1373.06	&	-6.46	\\ \hline
89/02/01	&	1366.1	&	1372.50	&	-6.40	&	&	89/11/08	&	1366.5	&	1372.67	&	-6.17	\\ \hline
89/02/12	&	1365.6	&	1371.89	&	-6.29	&	&	89/11/23	&	1365.7	&	1372.51	&	-6.81	\\ \hline
89/02/24	&	1365.5	&	1371.85	&	-6.35	&	&	89/11/30	&	1365.5	&	1372.65	&	-7.15	\\ \hline
89/03/01	&	1365.8	&	1372.25	&	-6.45	&	&	89/12/15	&	1366.4	&	1373.02	&	-6.62	\\ \hline
89/03/22	&	1365.3	&	1371.67	&	-6.37	&	&	89/12/20	&	1365.9	&	1369.71	&	-3.81$^*$	\\ \hline
89/03/29	&	1365.9	&	1372.38	&	-6.48	&	&	90/01/03	&	1366.1	&	1372.89	&	-6.79	\\ \hline
89/04/12	&	1366.8	&	1373.64	&	-6.84	&	&	90/01/17	&	1366.0	&	1372.88	&	-6.88	\\ \hline
89/04/26	&	1365.9	&	1372.12	&	-6.22	&	&	90/01/31	&	1366.3	&	1372.56	&	-6.26	\\ \hline
89/05/10	&	1366.7	&	1373.33	&	-6.63	&	&	90/02/11	&	1365.6	&	1372.47	&	-6.87	\\ \hline
89/05/24	&	1365.9	&	1371.86	&	-5.96	&	&	90/02/22	&	1365.5	&	1372.20	&	-6.70	\\ \hline
89/06/02	&	1365.6	&	1372.23	&	-6.63	&	&	90/02/28	&	1365.7	&	1372.19	&	-6.49	\\ \hline
89/06/15	&	1363.3	&	1369.69	&	-6.39	&	&	90/03/14	&	1366.2	&	1372.40	&	-6.20	\\ \hline
89/06/22	&	1366.1	&	1372.23	&	-6.13	&	&	90/03/28	&	1366.4	&	1372.83	&	-6.43	\\ \hline
89/07/06	&	1365.9	&	1372.72	&	-6.82	&	&	90/04/11	&	1365.9	&	1372.86	&	-6.96	\\ \hline
89/07/20	&	1366.1	&	1371.86	&	-5.76	&	&	90/04/25	&	1366.3	&	1373.21	&	-6.91	\\ \hline
89/08/03	&	1365.5	&	1371.72	&	-6.22	&	&	90/05/09	&	1365.9	&	1373.03	&	-7.13	\\ \hline
89/08/15	&	1365.7	&	1371.39	&	-5.69	&	&	90/05/24	&	1366.4	&	1372.56	&	-6.16	\\ \hline
89/08/25	&	1366.6	&	1372.90	&	-6.30	&	&	90/05/30	&	1365.6	&	1372.81	&	-7.21	\\ \hline
89/08/30	&	1366.2	&	1372.60	&	-6.40	&	&	90/06/14	&	1366.6	&	1373.34	&	-6.74	\\ \hline
89/09/13	&	1365.9	&	1371.62	&	-5.72	&	&	90/06/20	&	1366.3	&	1373.26	&	-6.96	\\ \hline
\end{tabular} \caption{TSI data from ERBS/ERBE and Nimbus7/ERB and their difference, which
is depicted in Figure 6. $^{*}$The value on December 20, 1989 is
excluded because the ERB value is highly uncertain. Data from \protect\url{ftp://ftp.ngdc.noaa.gov/STP/SOLAR\_{}DATA/SOLAR\_{}IRRADIANCE/} }
\end{table*}

\section{ACRIM - PMOD - KBS07 comparison}

\citet{ScafettaWillson2009} showed that the TSI proxy model of \citet{Krivova2007}
is not compatible with an ERB sensitivity increase in September 1989.
This was demonstrated by showing that bridging the ACRIM Gap using
KBS07 instead of ERB produces a 1980-2000 upward trend very similar
to that found by the ACRIM TSI composite. Here, we confirm and extend
this result with an alternative methodology.

In this section and in the following we study dynamical pattern divergences
between the TSI satellite records and TSI proxy models. If the function
$TSI_{sat}(t)$ represents a TSI experimental satellite record during
a given period, and the function $TSI_{mod}(t)$ represents a TSI
proxy model supposed to reconstruct the experimental record, then
the residual function

\begin{equation}
f(t)=TSI_{sat}(t)-TSI_{mod}(t)\label{eq:1}
\end{equation}
can be computed. Clearly if the proxy model well reproduces the experimental
result within the given time interval, then the function $f(t)$ should
be compatible with stationary random noise. However, if statistically
significant trends are observed in $f(t)$ the proxy model is not
capable of reproducing the experimental results. The use of this methodology
to study TSI records is common in scientific literature \citep[compare with: ][]{Frohlich2004,Frohlich2006,Frohlich2012}.

\begin{figure*}[!t]
\centering
\includegraphics[width=1.4\columnwidth, angle=-90]{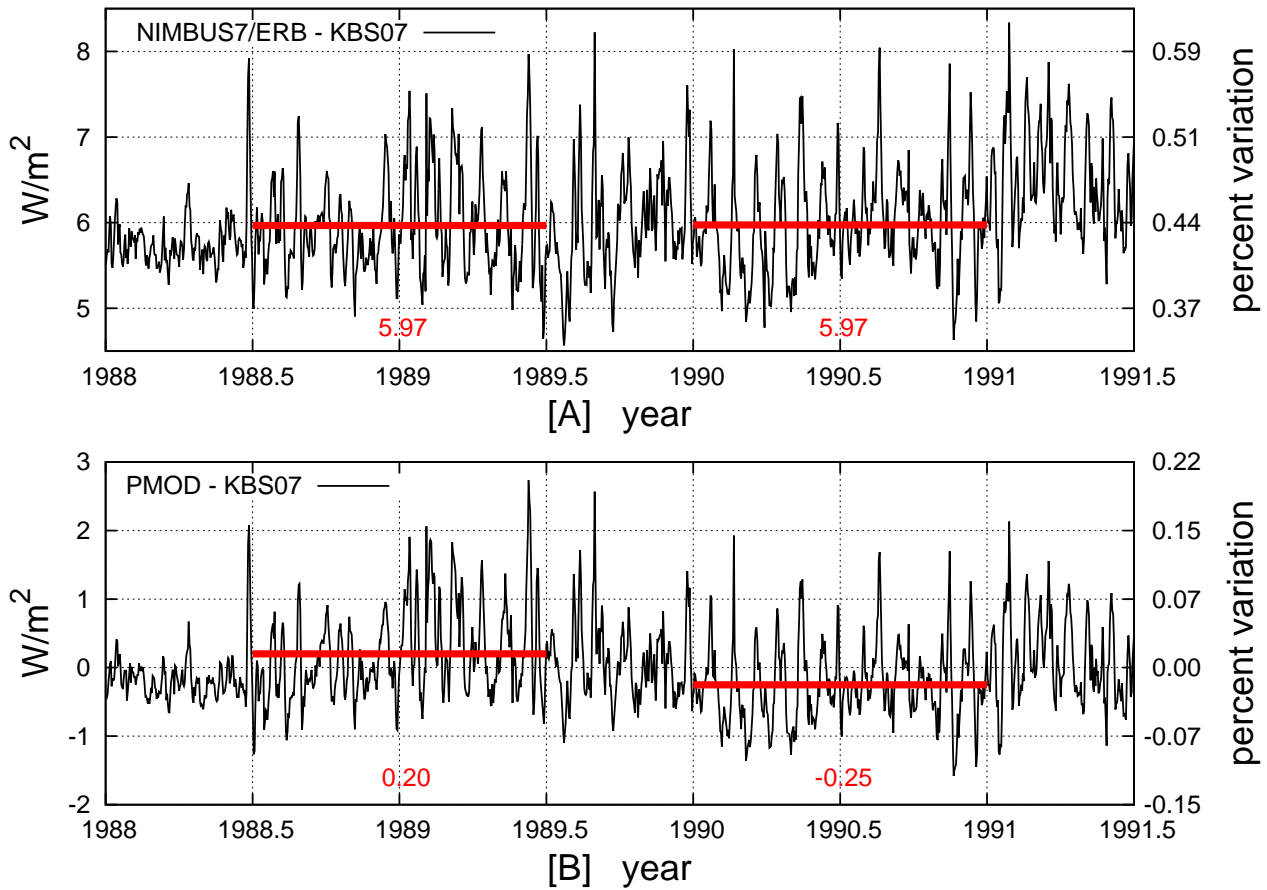}\caption{Residual between ERB, PMOD and the KBS07. {[}A{]} residual function
Eq. \ref{eq:1} between ERB results and the KBS07 proxy model proposed
by \citet{Krivova2007}. {[}B{]} residual function Eq. \ref{eq:1}
between the PMOD composite and KBS07. Red segments and red values
indicate the relative levels during the period 1988.5-1989.5 and 1990-1991
(before and after Fr\"ohlich's hypothesized ERB-glitch on 09/29/89).
The relative average values during 1988.5-1989.5 and 1990-1991 show
that KBS07 does not support the 09/29/89 ERB sensitivity step increase
proposed by Fr\"ohlich.}
\end{figure*}

Figure 7 shows the residual function (Eq. \ref{eq:1}) between ERB
and KBS07 (top panel) and between PMOD and KBS07 (bottom panel). Clearly
the KBS07 model shows the same trends as ERB both before and after
09/29/1989. However, KBS07 shows a significant shift of 0.45 $W/m^{2}$
relative to PMOD. This is caused by the putative PMOD sensitivity
change (+0.033\%) applied to the ERB results for that day. Thus, the
KBS07 model shows the same upward trend as the original ERB record
from 1988.5 to 1991. The ERB ACRIM Gap sensitivity increases proposed
by \citet{Lee}, \citet{Chapmam}, \citet{Frohlich1998}, \citet{Frohlich2004}
and \citet{Frohlich2006} are incompatible with the KBS07 proxy model.

\citet{Krivova2009} criticism of a previous analysis of the ACRIM
Gap by \citet{ScafettaWillson2009} claimed that the KBS07 model was
only useful for longer periods than the ACRIM Gap and lacked sufficient
resolution to reproduce trends as short as 1-2 year periods. To show
Krivova\textquoteright{}s claims are incorrect we employ an alternative
methodology that excludes the ACRIM Gap period and ACRIM1, ACRIM2
and PMOD are directly compared against KBS07 during the longer near
decadal length periods pre-gap: 1980-1989.5 and post-gap: 1992.5-2001.

\begin{figure*}[!t]
\centering
\includegraphics[width=1.4\columnwidth, angle=-90]{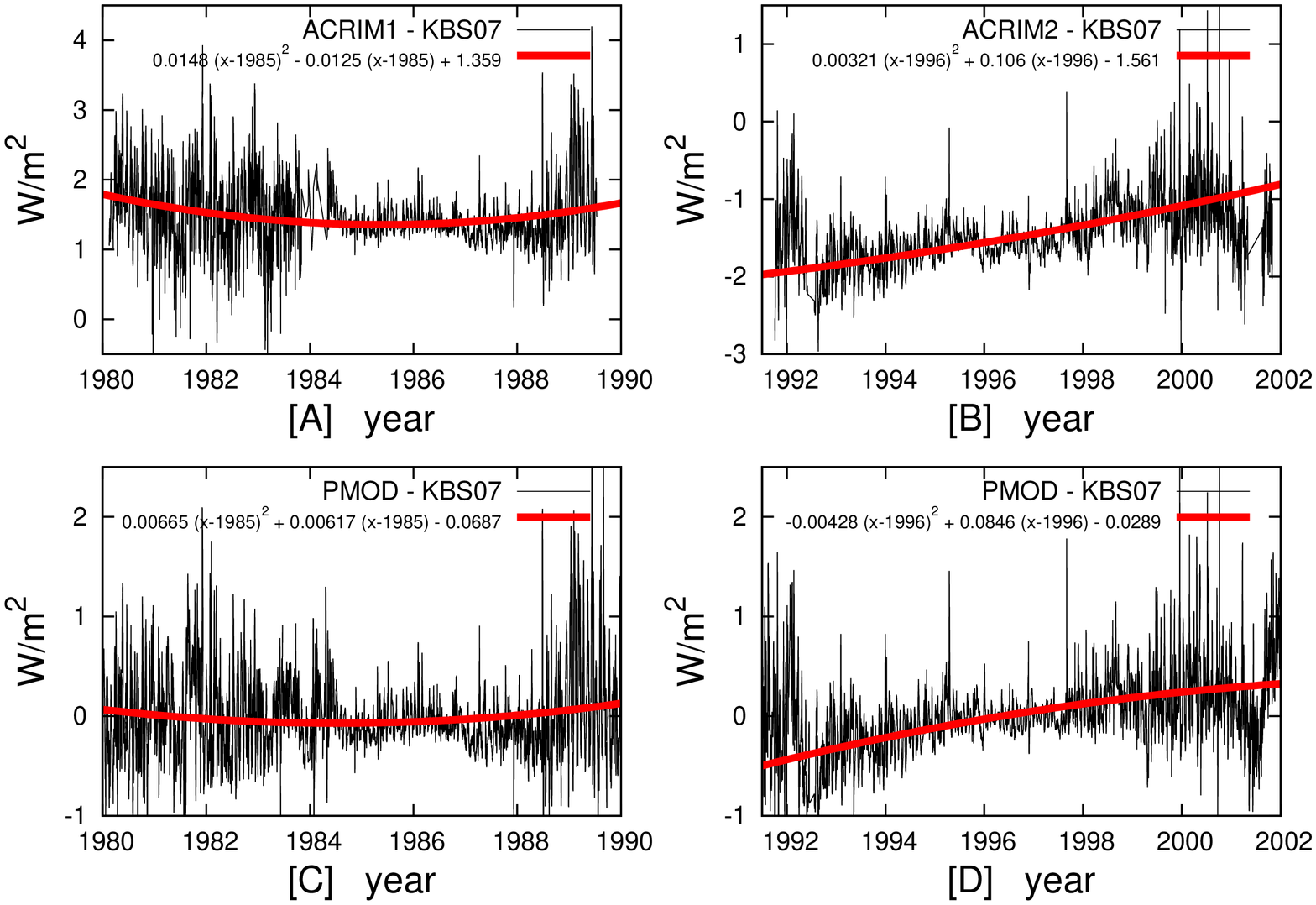}\caption{Residual function Eq. \ref{eq:1} of ACRIM1, ACRIM2 and PMOD with
the KBS07 TSI proxy model. The black curves are parabolic fits during
1980-1989.5 and 1992.5-2001.}
\end{figure*}

Figure 8 shows the residual function (Eq. \ref{eq:1}) between ACRIM1-ACRIM2
results and KBS07 (panels A and B), and between PMOD and KBS07 (panels
C and D). The results are evidently non stationary. KBS07 does not
capture the TSI decadal dynamics well. Second order polynomials (red
curves) are used to capture, at the first and second order of precision,
the discrepancies between experimental and the proxy model records\textquoteright{}
decadal trends and the curvature of the 11-year solar cycles. Figure
8 clearly shows that on the decadal scale KBS07 underestimates the
amplitude of the solar cycle between 1980 and 1989 (that is, the polynomial
fits of the residual functions present positive quadratic coefficients)
and misses an upward trend from 1992 to 2000 (that is, the polynomial
fits of the residual functions present positive linear coefficients).

\begin{figure*}[!t]
\centering
\includegraphics[width=1.6\columnwidth]{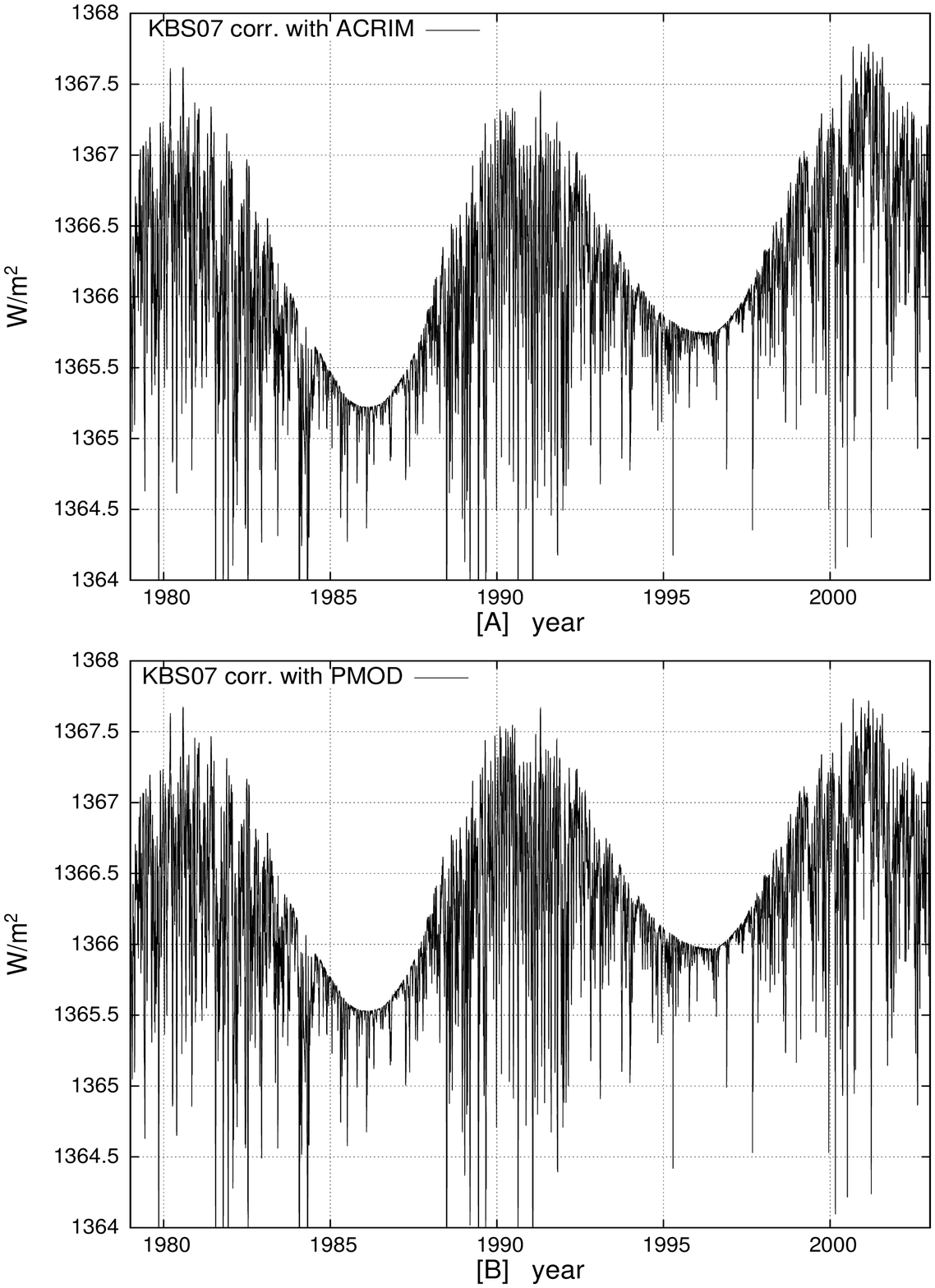}\caption{Top: KBS07 recalibrated on the ACRIM1 and ACRIM2 TSI records. Bottom:
KBS07 recalibrated on the PMOD TSI composite. The calibration covers
only the periods 1980-1989.5 and 1992.5-2001. The TSI 1980-2000 trends
of both are similar to the trend in the ACRIM composite.}
\end{figure*}

Figure 9 shows a KBS07 model after empirical adjustments during 1980-1989.5
and 1992.5-2001. The adjustments are designed to reproduce both the
trending and the amplitude of the 11-year solar cycles (as shown in
ACRIM1-ACRIM2 and PMOD TSI records) by using as a guide the second
order polynomial fit functions depicted in Figure 8. For example,
if $p(t)$ is the second order polynomial fit of the residual function
(Eq. \ref{eq:1}) shown in Figure 8 in a specific period 1980-1990
or 1991.5-2000, the KBS07 model is corrected as $KBS07+p(t)$. This
implies that: (1) the amplitude of the KBS07 11-year solar cycle from
1980 to 1989 KBS07 is corrected by increasing its amplitude by the
amount shown in Figure 8A and 8C, respectively, which lowers the TSI
minimum in 1986; and (2) from 1992 to 2000 KBS07 is corrected by adding
the upward trending by the amount shown in Figures 8B and 8D, respectively,
which raises the TSI minimum in 1996. The final result makes KBS07
better resemble the dynamical patterns of ACRIM1 and ACRIM2 records
(Figure 9A) and of the PMOD TSI composite (Figure 9B) outside the
ACRIM gap. The KBS07 data during the ACRIM Gap are left unaltered.
Figure 9 shows that once the KBS07 model is adjusted to better fit
the data, a TSI upward trend emerges in both KBS07-adjusted composites
during 1980-2000. This upward trend resembles the upward trend of
the original ACRIM TSI composite.

Thus, the multi-decadal agreement between the original KBS07 model
and PMOD derived by \citet{Krivova2007} appears coincidental. It\textquoteright{}s
likely an artifact of the failure of KBS07 to reproduce the correct
trending and amplitude of the solar cycles from 1980 to 2001. The
ACRIM1 and ACRIM2 results provide the best available estimates of
TSI during this period. This result supports the analysis of \citet{ScafettaWillson2009}
and validates the ACRIM TSI composite.

\section{ERB - ACRIM1 - PMOD - WFKS06 comparison}

We will now compare the TSI data against the WFKS06 solar magnetic
field strength models \citep{Wenzler2006,Wenzler2009}. \citet{Krivova2009}
claims that they are more accurate than KBS07 on short time scales
and agree with PMOD ACRIM gap hypothesis.

In addition to the ACRIM Gap ERB TSI data, Fr\"ohlich altered other
TSI published results for the ERB and ACRIM1 before 1985 \citep{Frohlich1998,Frohlich2004,Frohlich2006,Frohlich2012}:
(1) the early solar cycle 21 peak of ERB results were altered to agree
with Lean's TSI proxy model; (2) ACRIM1 results were altered to include
Fr\"ohlich's speculation about uncorrected degradation during its
first year of operation; (3) Fr\"ohlich used his altered version
of published ERB results instead of ACRIM1 during 1981-1984, claiming
that ACRIM1 results were compromised during the SMM spin mode. None
of these adjustments are supported by Fr\"ohlich with physical arguments,
algorithm changes or computations using original data and all are
disputed by the ERB science team (see the Hoyt's statement in the
Appendix) and the ACRIM1 science team (which is represented here by
ACRIM PI Willson).

\begin{figure*}[!t]
\centering
\includegraphics[width=1.4\columnwidth, angle=-90]{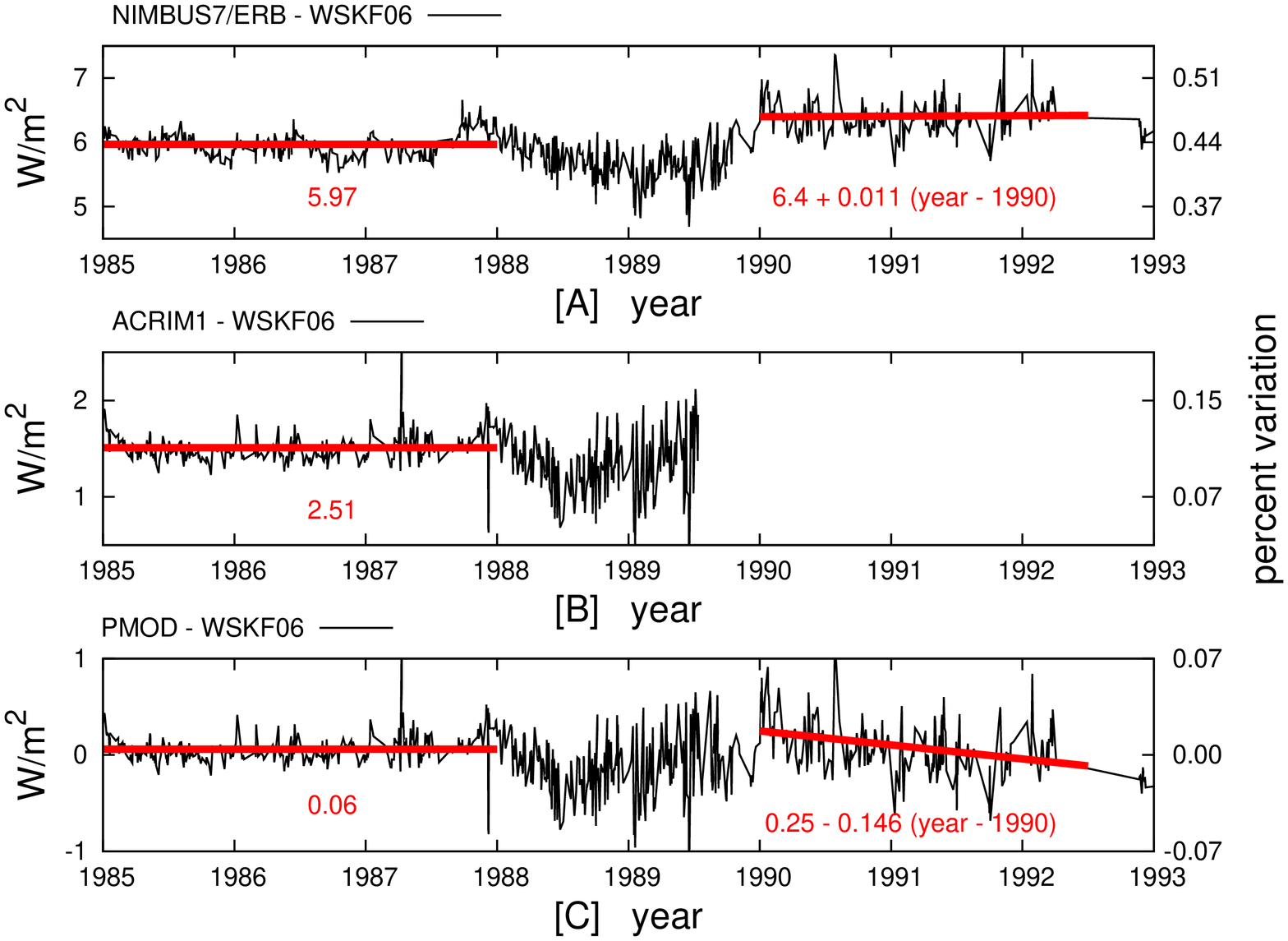}\caption{Residual function Eq. \ref{eq:1} between: {[}A{]} ERB and WSKF06;
{[}B{]} ACRIM1 and WSKF06; {[}B{]} PMOD and WSKF06. It is used the
WSKF06 optimum model \citep{Wenzler2006}. }
\end{figure*}

\begin{figure*}[!t]
\centering
\includegraphics[width=1.4\columnwidth, angle=-90]{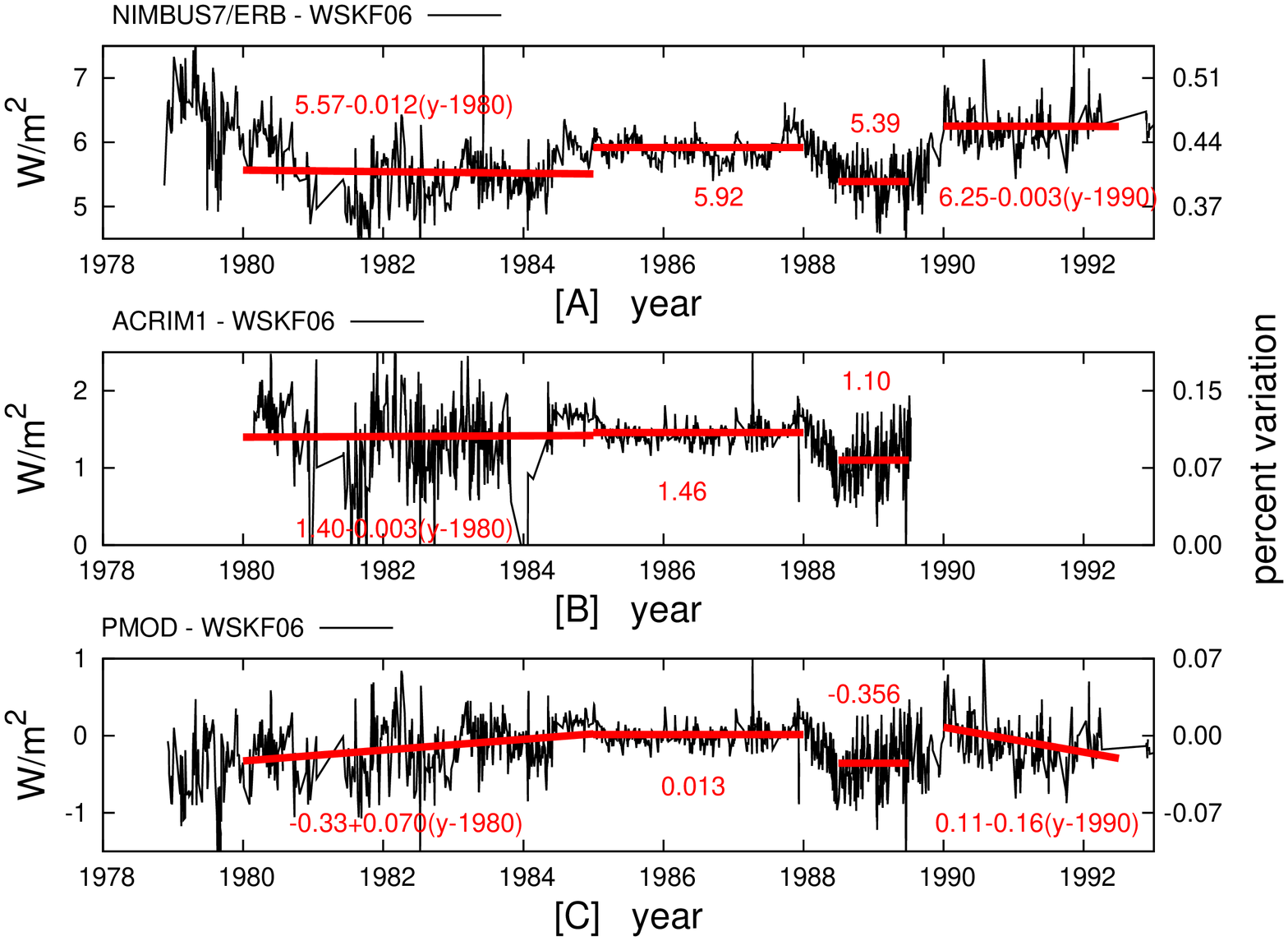}\caption{Residual function Eq. \ref{eq:1} between: {[}A{]} ERB and WSKF06;
{[}B{]} ACRIM1 and WSKF06; {[}C{]} PMOD and WSKF06. It is used the
WSKF06 standard model \citep{Wenzler2006}.}
\end{figure*}

WSKF06 TSI proxy reconstructions \citep{Wenzler2006} (see figure
3B and 3C) are based on two inhomogeneous records of the National
Solar Observatory (NSO): 1734 data points are from the 512-channel
Diode Array Magnetograph (NSO-512) which covers the period from Feb/1/74
to Apr/18/92 and from Nov/28/92 to Apr/10/93) and 2055 data points
from the Kitt Peak Spectromagnetograph (NSO-SPM) which covers the
period from Nov/21/92 to Sep/21/03). Only 45 days of NSO-512 data
(Nov/28/92 to Apr/10/93) are used for the cross calibration between
NSO-SPM and NSO-512 data. Thus, NSO-512 and NSO-SPM records were collected
using different instrumentation and their composite in 1992 includes
a significant cross-calibration uncertainty as a result.

\citet[figure 6]{Wenzler2006} shows the histogram-equating curves
for 22 individual days calculated with NSO-SPM and NSO-512 magnetograms
used to cross-calibrate the two records and construct the WSKF06 composite.
This figure clearly suggests a nonlinear relationship between these
two magnetogram records. Despite this \citet{Wenzler2006} simplified
their analysis by assuming linearity between the two data sets and
proposed two solutions: (1) a \textit{standard} model using the accepted
value of the cross-correlation factor $f=1.46$ between NSO-512 and
NSO-SPM and (2) an \textit{optimized} model using a cross-correlation
factor $f=1.63$ specifically chosen to reproduce a minimum to minimum
trend agreeing with the PMOD composite. ACRIM\textquoteright{}s trend
can also be reproduced by choosing a factor of $f=2.0$, which \citet{Wenzler2009}
rejected as a high-end value for $f$.

It is clear that using this model to discriminate between ACRIM and
PMOD trending between 1980 and 2000 could not produce unambiguous
results due mostly to the cross-calibration uncertainty problem between
NSO-512 and NSO-SPM. Our approach is to compare WSKF06 directly with
the satellite observations which adds information that may help identify
the correct TSI composite.

We make use of the residual function given by Eq. \ref{eq:1} between
ERB, ACRIM1 and PMOD results as shown in Figures 10 and 11. The predictions
of the WSKF06 \textit{optimum} and \textit{standard} proxy model results
are shown in Figures 10 and 11, respectively. During this period (prior
to 1992) WSKF06 is comprised of only the NSO-512 record, which removes
the cross-calibration uncertainty problem between NSO-512 and NSO-SPM
discussed above.

\subsection{Period 1978-1980: the Nimbus7/ERB 1979 peak}

Let us now analyze alternative periods. The original ERB results show
a peak during 1979.1-1979.3 that \citet{Frohlich1998} reduce in their
PMOD composite to agree with the prediction of Lean\textquoteright{}s
proxy model. Figure 12 compares the WSKF06 model and PMOD composite
during the period 1979.5 to 1979.6. WSKF06 presents a TSI peak in
1979.2 about $0.8$ $W/m^{2}$ higher than the PMOD level, although
its amplitude is lower than that of the original ERB record (compare
with Figure 11). Fr\"ohlich's reduction appears to be excessive since
WSKF06 in Figure 12 also shows a TSI peak in early 1979. It\textquoteright{}s
worthwhile mentioning that the ERB TSI peak in early 1979 is very
similar to a pronounced peak during the maximum of solar cycle 23
(1998-2004) that was observed by both ACRIM2, ACRIM3 and VIRGO (see
Figure 1) during comparable solar magnetic activity. Recently, \citet{ScfettaW2013b}
showed that these TSI peaks are correlated with the 1.092-year conjunction
cycle between Jupiter and Earth, fitting a pattern of planetary modulation
of solar activity.

\begin{figure*}[!t]
\centering
\includegraphics[angle=-90,width=1.6\columnwidth]{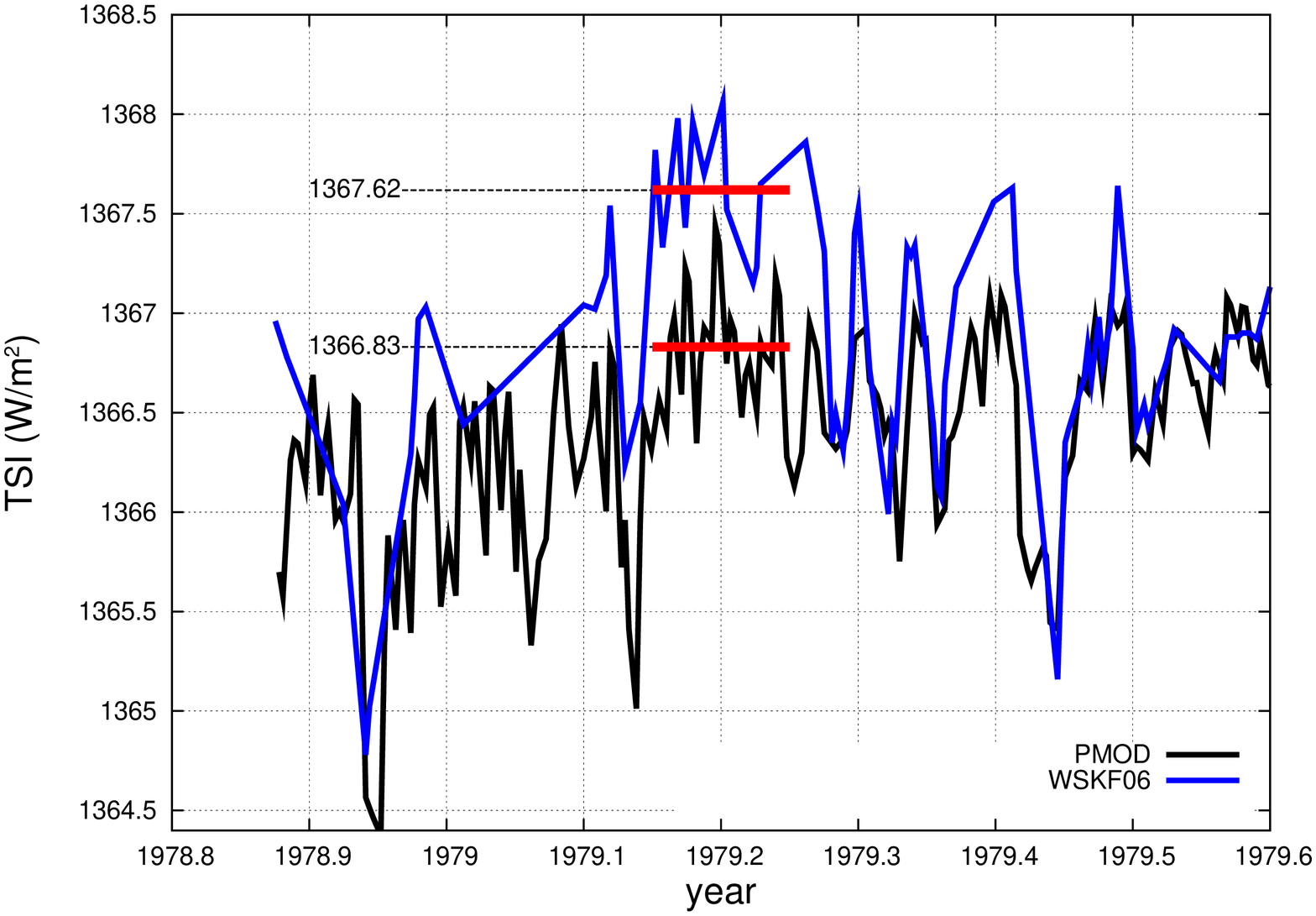}\caption{WSKF06 standard TSI proxy model against the PMOD TSI composite. The
large divergence between the two records during 1979.1-1979.3 corresponds
to the peak in ERB results that was mostly removed by the PMOD science
team to conform the TSI satellite records to the predictions of Lean's
proxy model.}
\end{figure*}

The comparison with WSKF06 indicates that before 1980 ERB results
can require some adjustment for possible uncorrected degradation and
a change of the orbital orientation as proposed by \citet{Frohlich2004,Frohlich2006}.
The adjustments to ERB results made by Fr\"ohlich in the PMOD composite
are too large since they remove a TSI peak near 1979.2 that is predicted
by WSKF06 and present in the original ERB results.

\subsection{Period 1980-1985: the ACRIM1 spin mode}

The trend agreement between ACRIM1 and the WSKF 06 proxy model during
1980 to 1985 is excellent, as seen in Figure 11. The linear fit of
the residual function (Eq. \ref{eq:1}) between the two records produces
a slope statistically equivalent to zero ($-0.003\pm0.02$ $Wm^{-2}/year$),
which indicates statistical stationarity. The comparison with ERB
also yields a slope statistically equivalent to zero ($-0.012\pm0.015$
$Wm^{-2}/year$). In contrast, a poor trend agreement is found between
PMOD and WSKF06 as measured by the statistically significant positive
slope ($+0.07\pm0.01$ $Wm^{-2}/year$). This is another counter-indication
for the PMOD corrections and Fr\"ohlich\textquoteright{}s choice
of altered ERB to fit Lean's TSI model \citep{Frohlich1998} rather
than the original ACRIM1 results during this period. The PMOD/WSKF06
disagreement in this period and the agreement between the PMOD composite
and Lean\textquoteright{}s proxy model \citep{Frohlich1998} is yet
another demonstration of the limitations of Lean\textquoteright{}s
proxy model for characterizing TSI during this period.

\subsection{Period 1985-1988: general agreement}

From 1985 to 1988 good agreement is observed between the WSKF06, ACRIM1
and ERB results. No trend is found in their residual function given
by Eq. \ref{eq:1}. Good agreement is also seen between WSKF06 and
PMOD and this is a direct result of PMOD\textquoteright{}s use of
unaltered ACRIM1 results during this period. Overall, Figure 11 shows
that from 1980 to 1988 there is better agreement between WSKF06 and
ACRIM1 than between WSKF06 and PMOD.

\subsection{Period 1988-1990: an upward shift of WSKF06}

Figures 10 and 11 clearly show that WSKF06 diverges significantly
from the results of ERB, ACRIM1 and the PMOD composite during 1988
to 1990. WSKF06 increases more rapidly than the TSI observations by
about 0.3-0.5 $W/m^{2}$ (see Figure 11 for details). This indicates
that WSKF06 does not reproduce TSI accurately during the ascending
phase of solar cycle 22, as acknowledged by \citet{Wenzler2006}.
This period and the one before 1988 are mismatched by about 0.5 $W/m^{2}$.

This is important because the \citet{Krivova2009} analysis disagreed
with the earlier results of \citet{ScafettaWillson2009} in which
a mixed ACRIM-KBS07 composite was constructed using KBS07 from 1988
to 1993 to bridge the ACRIM Gap and to merge ACRIM1 and ACRIM2. \citet{Krivova2009}
used the same methodology of \citet{ScafettaWillson2009} but substituted
the WSKF06 model for KBS07 and found a result different from ours.
The mistake of \citet{Krivova2009} was their failure to recognize
the drift of WSKF06 relative to ACRIM1 from 1988 to 1989 (see Figures
10 and 11), which clearly counter-indicates its use for merging ACRIM1
and ACRIM2. The 1988-1989 drift would cause an ACRIM-WSKF06 composite
to artificially shift the 1980-1989 ACRIM1 data $0.5$ $W/m^{2}$
upward relative to the 1992-2000 period. This artifact would reproduce
the 1980-2000 PMOD pattern and obscure the 1980-2000 TSI upward trend
that is common to both the ACRIM and ACRIM-KBS07 composites. Therefore,
the criticism of \citet{ScafettaWillson2009} by \citet{Krivova2009}
and their conclusions supporting the PMOD relative to the ACRIM composite
is not correct.

\subsection{Period 1990-1992.5: a Nimbus7/ERB upward drift? }

Figures 10 and 11 show stationarity in the residual function (Eq.
\ref{eq:1}). Both the standard and optimum WSKF06 models are stable
during the 1990-1992.5 ACRIM Gap period, and reproduce the trending
of unaltered ERB results very well. Linear fits of the ERB-WSKF06
differences show no significant trends during this period for either
the \textit{standard} model ($0.003\pm0.03$ $Wm^{-2}/year$) or the
\textit{optimized} model ($0.011\pm0.03$ $Wm^{-2}/year$). By contrast
the comparison of WSKF06 and PMOD during the same period reveals a
very different result. The PMOD-WSKF06 difference linear fits have
significant downward trending for both the standard and optimum WSKF06
models ($-0.16\pm0.03$ $Wm^{-2}/year$ and $-0.146\pm0.03$ $Wm^{-2}/year$
respectively). The PMOD/WSKF06 difference trends are comparable to
the correction Fr\"ohlich applied to ERB results during this period
(see Figure 4). The discrepancy between PMOD and WSKF06 is a direct
consequence of Fr\"ohlich\textquoteright{}s downward adjustment of
the ERB data.

The PMOD composite is based on a shift of ERB results totaling $\sim0.86$
$W/m^{2}$ during the ACRIM Gap, more than twice the $\sim0.4$ $W/m^{2}$
ERB shift derived from comparison with WSKF06 from 1990 to 1992. Therefore,
reconciliation of the PMOD composite to WSKF06 following the ACRIM
Gap would require an upward adjustment of $\sim0.46$ $W/m^{2}$ that
conforms the PMOD solar cycle 21\textendash{}22 minima trend to that
of the ACRIM composite.

\subsection{Period 1990-1992.5 using the upgraded SATIRE model.}

\citet{Ball} recently upgraded WSKF06. They show in their figures
7-10 that from 1990 to 1992.5 both the ACRIM TSI composite and the
magnetogram-based SATIRE model trend upward during the ACRIM Gap while
the PMOD trends downward. Figure 13 reproduces one of Ball's figures
where the upward trending of the unaltered ERB record, as used in
ACRIM, is approximately reproduced by the SATIRE model, while PMOD
slopes downward during the same period like ERBE.

\begin{figure*}[!t]
\centering
\includegraphics[width=1.4\columnwidth]{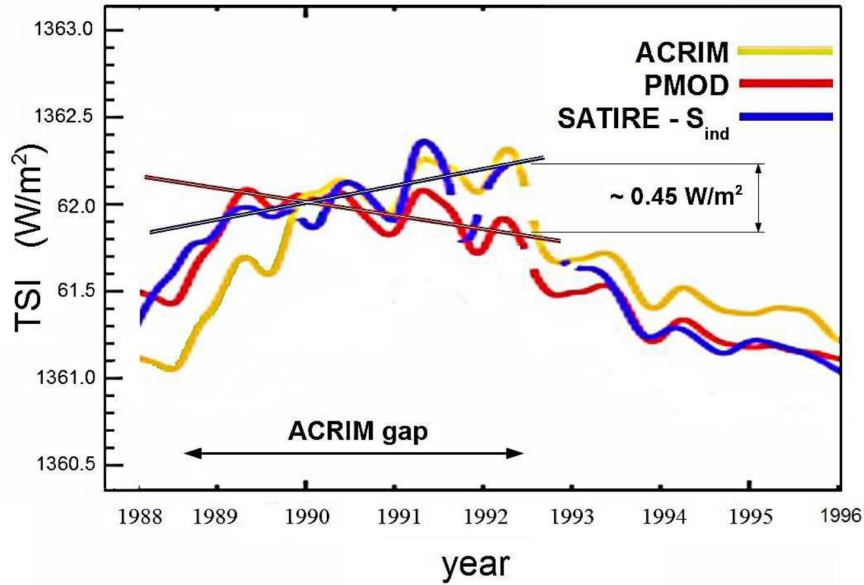}\caption{Comparison of smoothed ACRIM and PMOD TSI composites and the magnetogram-based
SATIRE-Sind TSI model during the ACRIM Gap \citep{Ball}. Note that
from 1990 to 1992.5 the SATIRE prediction trends upward approximately
like the ACRIM composite (slope = $0.1\pm0.03$ $Wm^{-2}/year$) while
the PMOD composite trends downward (slope = $-0.09\pm0.03$ $Wm^{-2}/year$).}
\end{figure*}

\subsection{Summary}

In summary, the direct ERB-WSKF06 and PMOD-WSKF06 comparisons depicted
in Figures 10, 11 and 13 during the ACRIM Gap support: (1) the correctness
of the originally published ERB observations (within error bounds)
that TSI increased during the ACRIM Gap; and (2) the hypothesis of
\citet{Willson2003} that the ERB/ERBE difference during the ACRIM
Gap is the result of ERBE degradation. Our results here directly contradict
the PMOD hypothesis of an ERB sensitivity drift during October 1989
through mid-1992.

\section{A close look at Lean's TSI proxy model}

The proxy model of \citet{Lean}, based on linear regression of sunspot
blocking and faculae brightening indexes against satellite TSI observations,
was first used to guide and validate the PMOD hypotheses \citep{Frohlich1998}.
As Figure 14 shows, Lean's TSI proxy model presents a slight TSI decrease
between the solar minima of 1986 and 1996, and a more evident TSI
decrease during the ACRIM Gap from 1989 to 1992. These patterns were
reproduced by PMOD by lowering the ERB results during the ACRIM Gap
to essentially agree with the ERBE data. This gave an impression of
mutual validation by Lean's model and Fr\"ohlich's \textquoteleft{}ACRIM
gap ERB glitch\textquoteright{} hypothesis. However, this is not the
case as we will now discuss.

\begin{figure*}[!t]
\centering
\includegraphics[angle=-90,width=1.4\columnwidth]{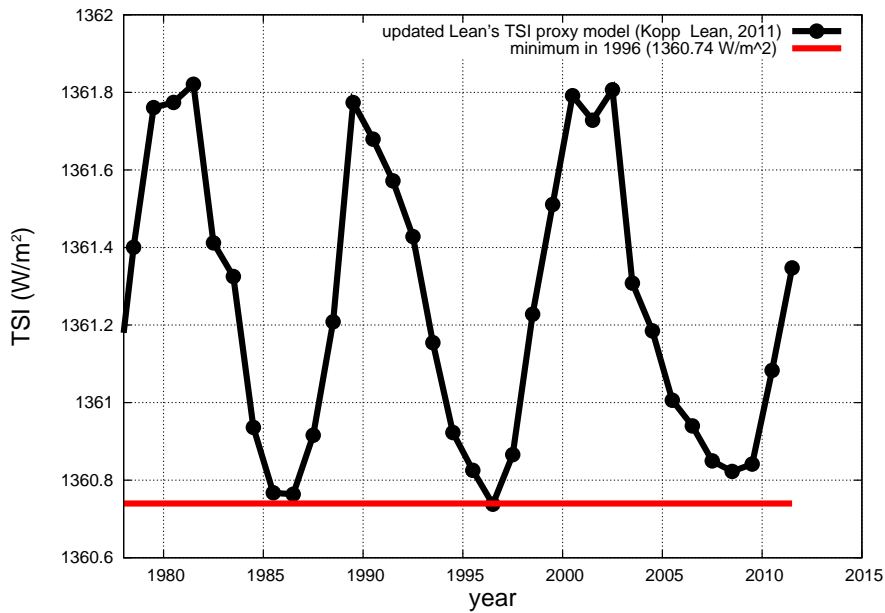}\caption{Updated Lean TSI proxy model \citep{Kopp2011}. Note that the TSI
minimum in 2008 is about 0.1 $W/m^{2}$ higher than the minimum in
1996, which is contradicted by both ACRIM and PMOD (see Figure 2).
Data from \protect\url{http://lasp.colorado.edu/sorce/tsi\_{}data/TSI\_{}TIM\_{}Reconstruction.txt}.}
\end{figure*}

Limitations of the predictive capability of Lean's updated model \citep{Kopp2011}
can be demonstrated using the data of the last solar cycle. Lean's
model predicts a 2008 TSI minimum higher than the minimum of 1996
by $\sim0.1$ $W/m^{2}$. However, this prediction is contradicted
by   both ACRIM and PMOD TSI composites that present the opposite
trend with the 2008 minimum about 0.2-0.3 $W/m^{2}$ \textit{lower}
than the minimum in 1996 (see Figure 2). Other indices of solar magnetic
activity (and TSI therefore), the open solar magnetic flux, the galactic
cosmic ray (GCR) flux and additional solar indices all show solar
activity higher in 1996 than in 2008 \citep{Lockwood2012}. This demonstrates
that on annual to decadal time scales Lean's model is affected by
a statistical uncertainty near $\pm0.5$ $W/m^{2}$. This is very
nearly equal to the divergence of the ACRIM and PMOD TSI composites
solar minima trends between 1986 and 1996 in determining (see Figure
2).

This discussion clearly indicates that use of Lean\textquoteright{}s
model as a guide to \textquoteleft{}re-evaluate\textquoteright{} published
satellite observations is not able to add useful information to the
understanding the TSI time series. Lean\textquoteright{}s model cannot
predict the decadal/multi-decadal trending of TSI with sufficiently
accuracy to discriminate between the ACRIM and PMOD TSI composite
trends. This applies in general and specifically to the validation
of the PMOD composite by Lean\textquoteright{}s model proposed for
the period 1978-1992 \citep{Frohlich1998}.

\section{Conclusion}

We have conducted several independent evaluations of the accuracy
of TSI satellite data and their composites. The ACRIM TSI composite
relies solely on the continuity of the results of overlapping satellite
experiments as understood and published by the flight experiment teams.
The ACRIM composite has a direct and exclusively experimental justification
\citep{Willson2003}. On the contrary, the PMOD TSI composite \citep{Frohlich1998,Frohlich2004,Frohlich2006,Frohlich2012}
is essentially a theoretical model originally designed to agree with
Lean\textquoteright{}s TSI proxy model \citep{Frohlich1998}. It relies
on postulated but experimentally unverified drifts in the ERB record
during the ACRIM Gap, and other alterations of the published ERB and
ACRIM results, that are not recognized by their original experimental
teams and have not been verified by the PMOD by original computations
using ERB or ACRIM1 data.

Our findings support the reliability of the ACRIM composite as the
most likely and precise representation of 35 years of TSI monitoring
by satellite experiments. The only caveat is that the ERB record prior
to 1980 may require some correction for degradation, but it would
be much less than used in the PMOD composite.

We argued that the ACRIM composite most closely represents true TSI
because the very corrections of the published TSI data made by Fr\"ohlich
to construct the PMOD composite are not supported by a direct comparison
between ERBE and ERB records in the proximity of September/October
1989.

Direct comparison of ERB and ERBE during 1989 showed that Fr\"ohlich\textquoteright{}s
postulated Sep/29/1989 step function increase of $0.4$ $W/m^{2}$
in ERB sensitivity, which coincided with a power down event, did not
occur. The KBS07 proxy model does not support Fr\"ohlich\textquoteright{}s
ERB \textquoteleft{}glitch\textquoteright{} either. A divergence between
the two satellite records did occur in November 1989; but this is
more than one month later and clearly not associated with the ERB
end-of-September power down event.

We have demonstrated that the update of Lean\textquoteright{}s TSI
proxy model \citep{Kopp2011}, used originally to validate PMOD\textquoteright{}s
lack of trending from 1980 to 2000 \citep{Frohlich1998}, has inadequate
predictive capability to properly reconstruct the TSI decadal trending.
Lean\textquoteright{}s model predicted an upward trend between the
TSI minima in 1996 and 2008 while both ACRIM and PMOD present a downward
trend. This demonstrates that Lean's proxy model cannot reconstruct
TSI decadal trending with a precision smaller than $\pm0.5$ $W/m^{2}$
the same order as the difference between PMOD and ACRIM TSI composites.
Thus, the use of Lean's TSI proxy model is not useful as a guide to
correct satellite measurements.

The WSKF06 TSI proxy models contradict the primary PMOD rationale
by the following findings: (1) there was a TSI peak in late 1978 and
early 1979 as recorded by ERB (although some early mission degradation
of the instrument may have been uncompensated for); (2) the ACRIM1
published record is more stable than ERB during 1980-1984 and should
be preferred for constructing a TSI composite during this period;
(3) ERB did not experience either the end-September 1989 step function
drift in sensitivity or the upward linear drift claimed by Fr\"ohlich
during 1990-1992.5. The latter result is also evident in the upgraded
SATIRE model \citep{Ball}. Thus, if ERB requires some correction
during the ACRIM Gap, our results suggest that Fr\"ohlich overestimated
those corrections by at least a factor of two due to the fact that
at least one of the two hypotheses (the ERB glitch in Sep/29/1989
or the ERB drift from Oct/1989 to 1992) are not confirmed by our cross-analysis.
The ERB-ERBE divergence during the ACRIM Gap most likely resulted
from uncorrected degradation of ERBE in its first exposure to short
wavelength fluxes driven by enhanced solar activity during the 1989-1993
solar maximum or other events. Consequently PMOD should be shifted
upward by about $0.5$ $W/m^{2}$ after 1992 which produces a 1980-2000
TSI upward trending similar to that observed in the ACRIM composite.

Our results demonstrated that the validity of TSI proxy models should
not be overestimated since they frequently produce conflicting results
and contradictory features. Although Solanki's TSI proxy models appear
to reproduce the lack of a trend during solar cycles 21 - 22 in the
PMOD TSI composite, they contradict one or more of the hypotheses
advocated by PMOD to alter the originally published TSI used in constructing
the PMOD TSI composite. Thus some of the arguments used to promote
the PMOD composite \citep{Frohlich1998,Wenzler2006,Krivova2007,Wenzler2009}
are little more than speculations and coincidences. Moreover, Lean\textquoteright{}s
and Solanki\textquoteright{}s TSI models differ significantly from
the TSI model proposed by \citet{Hoy1993} who constructed a TSI record
since 1700 using five alternative solar irradiance proxy indexes \textemdash{}
sunspot cycle amplitude, sunspot cycle length, solar equatorial rotation
rate, fraction of penumbral spots, and the decay rate of the sun spot
cycle.

Although Lean\textquoteright{}s and Solanki\textquoteright{}s models
present no TSI trend during 1980-2000, as shown in the PMOD composite,
there are other studies suggesting generally increasing TSI from 1970
to 2000. \citet{Shapiro} found a small increasing trend across the
1975, 1986 and 1996 solar minima followed by a decrease in the minimum
of 2008. Also the cosmic ray flux index would suggest a solar activity
increase from 1980 to 1996 \citep[figure 20]{scafetta2013c}. The
TSI pattern revealed in the ACRIM satellite composite is consistent
with a quasi 60-year solar cycle modulation, which appears to be one
of the major harmonic constituents of solar activity and should have
theoretically peaked around 2000 \citep{Ogurtsov,Scafetta2012a,Scafetta2013a,Scafetta2013EE}.
We conclude that solar activity may have presented a larger secular
variability and specific geometrical patterns that are quite different
from the Lean TSI model currently used to force the CMIP5 models.

\subsection*{Acknowledgment: }

The National Aeronautics and Space Administration supported Dr. Willson
under contracts NNG004HZ42C at Columbia University and Subcontracts
1345042 and 1405003 at the Jet Propulsion Laboratory.

\section*{Appendix A: Hoyt's statement about Nimbus7/ERB}

In 2008 Scafetta asked Hoyt to comment the alterations of the ERB
data implemented by Fr\"ohlich to produce the PMOD composite. Hoyt
returned by email the following statement where ``N7'' is for the
Nimbus7/ERB TSI record prepared by Hoyt and collaborators:

\textit{September 16, 2008. }

\textit{Dear Dr. Scafetta: Concerning the supposed increase in N7
sensitivity at the end of September 1989 and other matters as proposed
by Fr\"ohlich\textquoteright{}s PMOD TSI composite:}

\textit{1. There is no known physical change in the electrically calibrated
N7 radiometer or its electronics that could have caused it to become
more sensitive. At least neither Lee Kyle nor I could never imagine
how such a thing could happen and no one else has ever come up with
a physical theory for the instrument that could cause it to become
more sensitive.}

\textit{2. The N7 radiometer was calibrated electrically every 12
days. The calibrations before and after the September shutdown gave
no indication of any change in the sensitivity of the radiometer.
Thus, when Bob Lee of the ERBS team originally claimed there was a
change in N7 sensitivity, we examined the issue and concluded there
was no internal evidence in the N7 records to warrant the correction
that he was proposing. Since the result was a null one, no publication
was thought necessary.}

\textit{3. Thus, Fr\"ohlich\textquoteright{}s PMOD TSI composite
is not consistent with the internal data or physics of the N7 cavity
radiometer.}

\textit{4. The correction of the N7 TSI values for 1979-1980 proposed
by Fr\"ohlich is also puzzling. The raw data was run through the
same algorithm for these early years and the subsequent years and
there is no justification for Fr\"ohlich\textquoteright{}s adjustment
in my opinion.}

\textit{Sincerely, Douglas Hoyt }

\section*{Appendix B: The importance of the TSI satellite debate for solar
physics and climate change}

The Sun is a variable star \citep{Brekke}. However, the multi-decadal
trending of solar activity is currently poorly modeled and numerous
alternative proxy reconstructions have been proposed. Understanding
the correct amplitude and dynamics of solar variability is important
both for solar physics and climate change science.

The multi-decadal trending difference between the ACRIM \citep{Willson2003}
and PMOD TSI composites \citep{Frohlich1998,Frohlich2006} shown in
Figure 2 is important for understanding the multi-decadal variation
of solar dynamics and therefore for discriminating among solar models
used also to interpret climate changes. Because the ACRIM TSI composite
shows an evident upward pattern from 1980 to 2000 while PMOD shows
a slight downward trend during the same period, the former would suggest
a larger TSI low-frequency variability than the latter and different
TSI multidecadal variation mechanisms. The origin of a slowly varying
irradiance component may derive from changes in the solar faculae
and/or in the background solar radiation from solar quiet regions.
These mechanisms are currently poorly understood and modeled. However,
if TSI increased from 1980 to 2000, total solar and heliospheric activity
could have increased as well contributing significantly to the global
warming observed from 1980 to 2000 \citep{scafetta2005,scafetta2007,Scafetta2009,Scafetta2011,scafetta2012aa,Scafetta2013EE,scafetta2013c}.

The Coupled Model Intercomparison Project Phase 5 (CMIP5) used to
study climate change \citep{scafetta2013c} currently recommends the
use of a solar forcing function deduced from the TSI proxy model originally
proposed by Lean and collaborators \citep{Wang,Kopp}. Lean's recent
models show a relatively small secular trend (about 1 $W/m^{2}$)
from the Maunder minimum (1645-1715) to the present with a peak about
1960 and it is quasi stationary since. Alternative TSI proxy reconstructions
have been proposed and some of them present much larger secular variability
and different decadal patterns. Figure 15A depicts two of these sequences:
Lean's TSI model and the TSI reconstruction proposed by \citet{Hoy1993}
rescaled at the ACRIM TSI level. Figure 15A also shows in blue the
annual mean ACRIM TSI satellite composite since 1981 \citep{Willson2003}.

\begin{figure*}[!t]
\centering
\includegraphics[width=1.6\columnwidth]{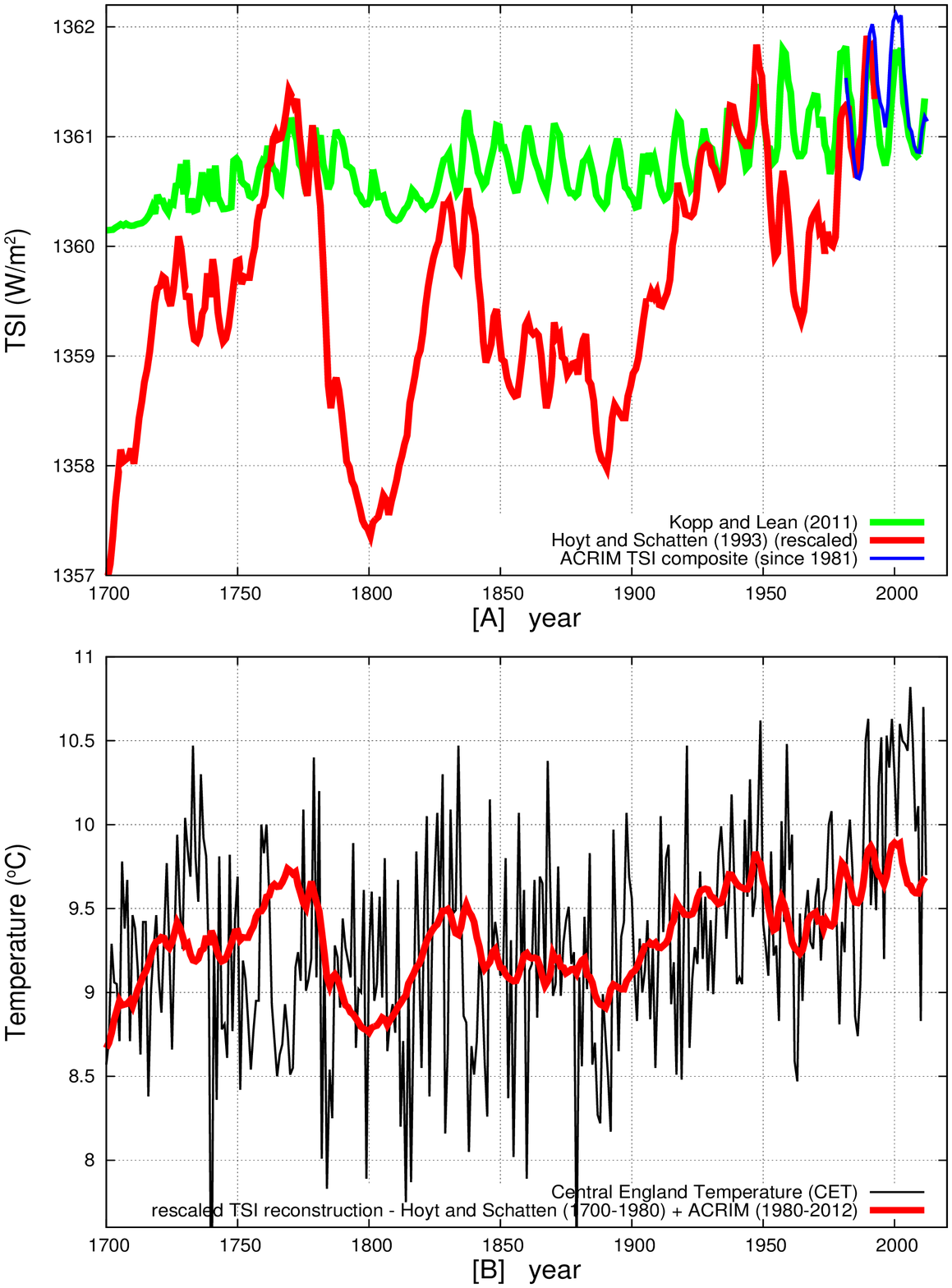}\caption{{[}A{]} Total solar irradiance (TSI) reconstruction by \citet{Hoy1993}
(red) rescaled on the ACRIM record \citep{Willson2003} (since 1981)
(blue) vs. the updated Lean model \citep{Wang,Kopp} (green). {[}B{]}
Comparison between the Central England Temperature (CET) record (black)
\citet{Parker} and the TSI model by Hoyt and Schatten plus the ACRIM
TSI record. Good correlation is observed at least since 1772. (Note
CET is less certain before 1772). The \citet{Hoy1993} reconstruction
has been made by rescaling it on the ACRIM record from 1980 to 1992
using the formula $HS93*1361.267/1371.844$, where $1371.844$ is
the 1981-1992 average of \citet{Hoy1993}'s proxy reconstruction and
$1361.267$ is the 1981-1992 average of the ACRIM TSI composite. The
value in 1980 in {[}B{]} was estimated as the average between the
ACRIM mean and the rescaled \citet{Hoy1993} reconstruction. }
\end{figure*}

\citet[fig. 10]{Hoy1993} showed that their multi-proxy TSI model
is highly correlated with an annual mean northern hemisphere temperature
variation reconstruction since 1700. This correlation is confirmed
(Figure 15B) by comparing a Hoyt+ACRIM TSI combination model against
the Central England Temperature record since 1700 \citep{Parker}.
The divergence observed during the last decades is likely due to an
additional anthropogenic warming component which was far weaker in
the past, as more clearly explained in the literature interpreting
global climate change \citep[e.g.:][]{scafetta2005,scafetta2007,Scafetta2009,Scafetta2010,Scafetta2011,scafetta2012aa,Scafetta2013EE,scafetta2013c}.
It has been demonstrated a good correlation between the same TSI proxy
model and numerous climatic records for the 20th century including
temperature records of the Arctic and of China, the sunshine duration
record of Japan and the Equator-to-Pole (Arctic) temperature gradient
record \citep{Soon2005,Soon,Soon2013}. Key features are a warming
from 1910s to 1940s, a cooling from the 1940s to 1970s, a warming
from the 1970s to 2000s and a steady-to-cooling temperature since
\textasciitilde{}2000, all of which correlate much better with the
Hoyt+ACRIM TSI composite than with Lean's proxy model.

Recently, \citet[see also the supplementary information]{Liu} used
the ECHO-G model and showed that to reproduce the $\sim0.7$ $^{o}C$
global cooling observed from the Medieval Warm Period (MWP: 900-1300)
to the Little Ice Age (LIA: 1400-1800) according to recent paleoclimatic
temperature reconstructions \citep[e.g.: ][]{Ljungqvist,Mann2008,Moberg},
a TSI model with a secular variability $\sim3.5$ times larger than
that shown by Lean\textquoteright{}s TSI model would be required.

The \citet[section 6.6.3.4 and its figure 6.14]{IPCC} reports that
to obtain a cooling of about 0.7 $^{o}C$ from the MWP to the LIA
Maunder Minimum a corresponding TSI downward trend of $-0.25\%$ is
required. Lean's TSI model shows a trend of only $-0.08\%$ over this
period \citep{Wang}. The same climate models rescaled using Lean's
TSI model predict a MWP-to-LIA Maunder Minimum cooling of only 0.25
$^{o}C$ that is compatible only with the controversial hockey stick
temperature reconstruction of \citet{Mann1999}. It should be noted
that the updated proxy temperature reconstructions by \citet{Mann2008}
show a significantly warmer MWP than the Mann's 1999 temperature reconstruction
used by the IPCC in 2001.

Thus, recent paleoclimatic temperature reconstructions imply that
the natural climate variability varied significantly more than predicted
by the CMIP5 general circulation models, which use Lean's low-variability
TSI model \citep[e.g.:][]{Scafetta2013b,Scafetta2013EE,scafetta2013c}.
The most likely explanation is that solar variations are a more significant
contributor to climate change than currently understood \citep[see also: ][]{Liu,Scafetta2013b}.
A stronger solar effect on the climate would also imply a significantly
larger solar contribution to the 20th century global warming, as demonstrated
in some works \citep{Scafetta2009,Scafetta2013b,Scafetta2013EE,scafetta2013c}.
Indeed, despite the \citet{IPCC} claims the sun has an almost negligible
effect on climate, numerous authors found significant correlations
between specific solar models and temperature records suggesting a
strong climate sensitivity to solar variations \citep[e.g.:][]{Bond,Hoy1993,Loehle,Mazzarella,Ogurtsov,Scafetta2009,Scafetta2010,Scafetta2012a,Scafetta2013EE,Schulz,Soon2005,Soon2013,Steinhilber,Svensmark,Thejll}.

Recently, \citet{Shapiro} and \citet{Judge} also proposed TSI models
based a comparison between solar irradiance reconstructions and sun-like-stellar
data that show a TSI secular variability at least 3-to-6 times greater
than Lean's TSI proxy, similar to those proposed by \citet{Hoy1993}.
The Shapiro model also predicts a small TSI increase between the solar
minima of 1986 and 1996, that is more consistent with the ACRIM 1980-2000
upward TSI pattern and contradicts PMOD. This pattern derives from
the fact that the cosmic ray flux record, which is inversely proportional
to solar magnetic activity, presents a slight decrease from about
1970 to 2000 \citep[figure 20]{scafetta2013c}.

It was recently speculated that long term changes in the solar interior
due to planetary gravitational perturbations may produce gradual multi-decadal
and secular irradiance changes \citep[e.g.:][]{Abreu,Charbonneau,Scafetta2012a,Scafetta2012b,Scafetta2013a}.
The planetary models proposed by \citet{Scafetta2012a} and \citet{Scafetta2013a}
shows a quasi 60-year modulation of solar activity since 1850 with
peaks in the 1880s, 1940s and 2000s. Thus, it shows good agreement
with the ACRIM composite\textquoteright{}s upward trending from about
1980 to 2000.

In conclusion, despite recent scientific climate change literature
\citep[e.g.: ][]{IPCC} has favored the PMOD interpretation of the
TSI experimental records we have provided experimental and theoretical
reasons for our belief that the ACRIM TSI composite is a most likely
interpretation of the current satellite TSI database. The dynamical
pattern revealed by the ACRIM TSI composite appears to better agree
with a number of new evidences that are emerging and, therefore, solving
the TSI satellite controversies could be quite important for better
understanding solar physics and climate change alike.

\end{document}